\documentclass{article}

\usepackage{arxiv}
\usepackage[dvipsnames]{xcolor}
 
\usepackage[T1]{fontenc}
\usepackage{dfadobe}  
\usepackage{todonotes}
\usepackage{tabu}
\usepackage{graphicx}
\usepackage{makecell}
\usepackage{array}
\usepackage{comment}
\usepackage{amsmath}
\usepackage{amsfonts}
\usepackage{gensymb}

\graphicspath{{./figs/}}

\title{QuickCurve: revisiting slightly non-planar 3D printing}

\usepackage{hyperref}

\usepackage{authblk}

\renewcommand{\shorttitle}{}

\author[1]{
Emilio Ottonello\thanks{\texttt{emilio.ottonello@iit.it}}%
}
\author[2]{
Pierre-Alexandre Hugron%
}
\author[1]{
Alberto Parmiggiani%
}
\author[2]{
\\Sylvain Lefebvre\thanks{\texttt{sylvain.lefebvre@inria.fr}}%
}

\affil[1]{Istituto Italiano di Tecnologia, Genova, Italy}
\affil[2]{Université de Lorraine, CNRS, Inria, LORIA, Nancy, France}

\begin{document}
\date{}
\maketitle
\begin{abstract}
Additive manufacturing builds physical objects by accumulating layers upon layers of solidified material.
This process is typically done with horizontal planar layers. However, fused filament
printers have the capability to extrude material along 3D curves. The idea of depositing out-of-plane, also known as non-planar printing, has spawned a trend of research towards algorithms that could generate non-planar deposition paths automatically from a 3D object.

In this paper we introduce a novel algorithm for this purpose. Our method optimizes for a curved slicing surface. This surface is intersected with the input model to extract non-planar layers, with the objective of accurately reproducing the model top surfaces while avoiding collisions. Our formulation leads to a simple and efficient approach that only requires solving for a single least-square problem. Notably, it does not require a tetrahedralization of the input or iterative solver passes, while being more general than simpler approaches. 
We further explore how to orient the paths to follow the principal curvatures of the surfaces, how to filter spurious tiny features damaging the results, and how to achieve a good compromise of mixing planar and non-planar strategies within the same part.

We present a complete formulation and its implementation, and demonstrate our method on a variety of 3D printed models.

\end{abstract}

\section{Introduction}

Additive manufacturing allows to fabricate 3D models by stacking flat layers of solidified material on top of each other. It enables the production of complex parts without the intrinsic limitations related to traditional machining. 

However, decomposing a part in flat layers leads to the so-called \textit{staircase defect} that damages the surface finish and accuracy of the final parts. This is evident on objects 3D printed with \textit{Fused Filament Fabrication (FFF)}, especially along slightly curved surfaces. Yet, FFF has the ability to deposit filament along 3D curves, fully exploiting 3-axis motion during deposition. This led to a line of research in \textit{non-planar printing}, which we review in Section~\ref{sec:prevwork}. By producing non--planar layers, these techniques attempt to improve the part in terms of its surface finish. Some techniques go beyond and propose to exploit additional degrees of freedom on robotic platforms. 

In this paper we focus on non-planar 3D printing on standard 3-axis machines, with the primary objective of improving the surface finish of the top surfaces of a print. The advantage of 3-axis methods is to be readily applicable on widely available FFF printers with little to no hardware modifications.

Non-planar printing on 3-axis machines is made difficult by the inability to orient the nozzle with respect to the slope. This leads to a nozzle \textit{gouging issue}, where the nozzle scratches already deposited material due to its physical shape. This limits the maximum angle that can be reproduced with satisfactory quality~\cite{ahlers_19_3d}. However, it is worth noting that the staircase defect \textit{worsens at lower slopes}. Therefore even a limited non-planarity brings great benefits in accuracy~\cite{etienne_19_curvislicer,ahlers_19_3d}.

Another important consequence of a fixed nozzle relates to the orientation of the paths with respect to the surface curvatures. There is a tradeoff in how the slope will be reproduced depending on how the paths align with the principal curvatures. If the paths align along the maximum curvature, their slopes follow the surface accurately. However, as the nozzle cannot incline significant gouging will occur. On the contrary if they align with the minimum curvature, less gouging occurs but the paths produce a staircase of their own, having their width across the slope. 
This issue is recognized in prior works~\cite{song_2016,ahlers_19_3d}, but to the best of our knowledge has never been directly addressed. We propose a novel strategy, orienting the paths according to the principal curvatures and explore the different tradeoffs.

Finally, we place a particular emphasis on designing an efficiency solver. The most advanced prior works require computationally expensive optimizations of tetrahedral meshes, under inequality constraints. In contrast, our technique requires a single least-square optimization pass. 

\begin{figure*}[tbh]
 \includegraphics[width=0.7\linewidth]{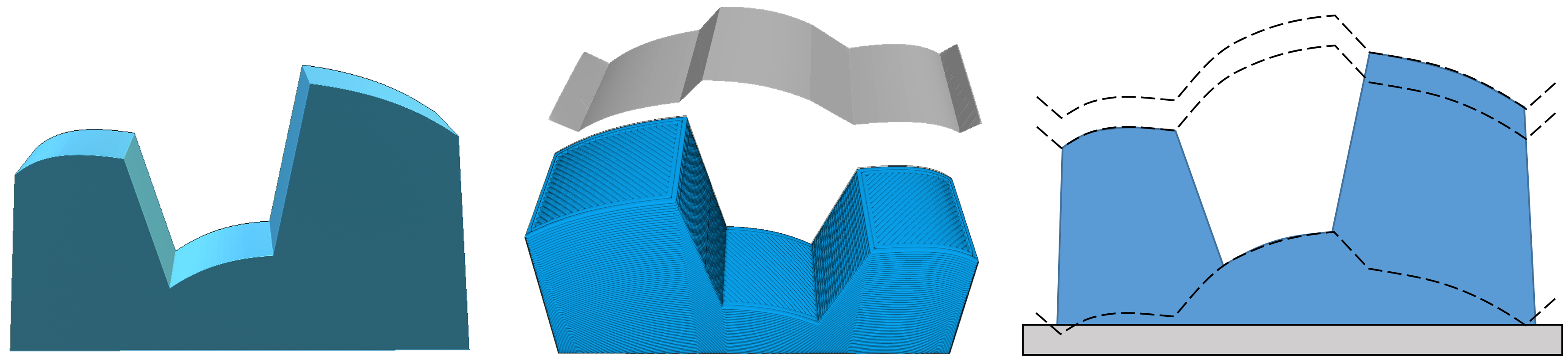}
 \includegraphics[width=0.29\linewidth]{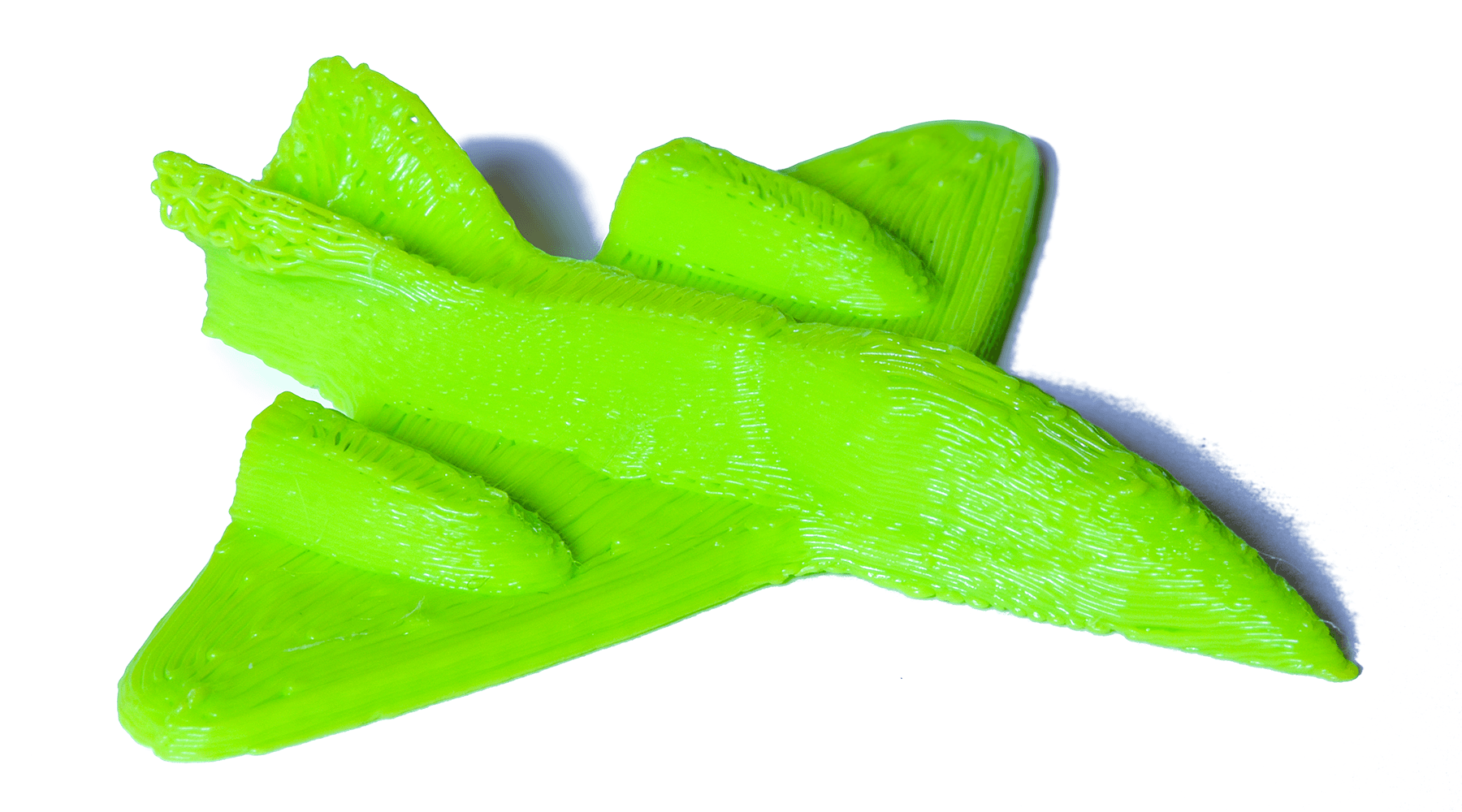}
 \centering
 \caption{Our technique starts from a 3D model to be 3D printed (left) using non-planar deposition to reproduce the top surface accurately (second, bottom). Our technique optimizes for a slicing surface (second, top) that is used to extract non-planar layers (third). The rightmost figure illustrates how the optimized slicing surface (dashed line) will precisely align with the top surfaces, extracting a layer that follows the curvature. We further propose a novel strategy capable of aligning the toolpaths along the curvature. Our approach allows  complex parts to be printed with a non-planar surface finish (rightmost). (Zoom in for details).}
\label{fig:teaser}
\end{figure*}

\noindent \paragraph*{Contributions} We propose:
\begin{itemize}
\item An efficient slicing algorithm optimizing for a non-planar slicing surface. The slicing surface geometry captures the top surfaces of a model accurately and ensures that no collisions can occur during fabrication. The algorithm is computationally efficient and works on the unmodified input model.
\item A toolpath algorithm that orients the paths of the top surfaces according to the surface curvature. This helps improve surface finish by aligning the deposition paths with either the maximum or minimum curvature.
\item An optional filtering approach that removes tiny features that could otherwise lead to spurious contours along the smooth, curved surfaces. This filter is especially effective on natural and noisy scanned surfaces, e.g. terrains.
\end{itemize}

All together, these contributions enable non-planar printing at a low computational overhead, while making it possible to fabricate parts with a good tradeoff in quality, with a control on how the part is decomposed in curved and sliced surfaces.







\section{Related work}
\label{sec:prevwork}

In this section, we review works relating to non-planar printing that are most closely related to ours.



\subsection{Non-planar printing on 3-axis machines}

Early experiments~\cite{Chakraborty_2007,allen_15_experimental} demonstrated how 3-axis printers can reproduce curved surface using commercial 3-axis printers. This was followed by the exploration of algorithms to automatically generate non-planar layers:

Song et al.~\cite{song_2016} propose to vary the toolpath thicknesses (height) on top layers, moving their top surfaces up and down to reproduce the true surface position -- an operation akin to antialiasing in Computer Graphics. This requires splitting and ordering the paths to minimize nozzle gouging, as well as careful flow management as the deposition thicknesses vary along each path.

Ezair et al.~\cite{Ezair2018} extract layers as iso-surfaces  of a user provided tri-variate parameterization, ensuring an equal distance between them. A similar technique extracts a dense set of toolpaths in each layer. Collisions are resolved as a postprocess, splitting and ordering toolpaths.

\begin{figure}
\centering
\includegraphics[width=0.4\linewidth]{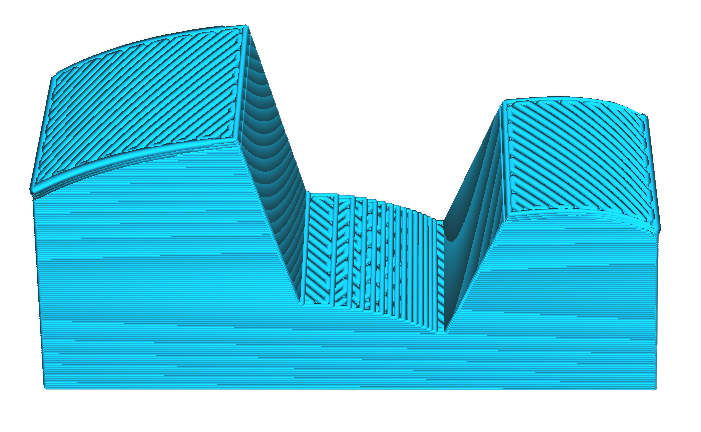}
\caption{The two-towers model. The approach of Ahlers et al.~\cite{ahlers_19_3d} used in this Figure cannot curve the middle part as it would produce collisions during printing. The result of our approach follows all surfaces, with curved layers throughout the part as shown in Figure~\ref{fig:teaser}. 
}
\label{fig:two_towers}
\end{figure}

Ahlers et al.~\cite{ahlers_19_3d} developed an algorithm to curve the accessible top layers of a print, while the part interior remains printed in a standard fashion. Curving only the top layers requires careful stitching with the surrounding flat layers. Surfaces that could be curved but would result in collisions during printing are excluded. This becomes problematic when top surfaces are at altitudes where one makes another inaccessible, see Figure~\ref{fig:two_towers} for an example.

%

The approach of Etienne et al.~\cite{etienne_19_curvislicer} optimizes for a {continuous} deformation of the entire input model \textit{volume}, with the objective of flattening top surfaces and steepening others. Flattening leads to curved surfaces while steepening improves the slicing quality of other surfaces, producing adaptive slicing as a side effect. The deformed model is sliced with flat layers, and the obtained toolpaths are deformed back, producing non-planar toolpaths of varying thicknesses throughout the part. The deformation is constrained to enforce (optional) minimal and maximal thicknesses, as well as avoid collisions: Once the mapping obtained the layers are guaranteed to be globally collision-free, requiring no explicit collision checks. 
While the method is very general and can curve surfaces throughout the part, it suffer a major downside: a discretization of the entire build volume by tetrahedrization. This makes the overall approach computationally heavy, especially as it iterates solving a large least-square problem under linear inequality constraints.

These techniques introduce important ideas for our purpose. A conical nozzle collision model~\cite{Ezair2018,ahlers_19_3d,etienne_19_curvislicer} leads to the observation that surfaces with an angle everywhere below a critical angle can be printed without self-collisions~\cite{ahlers_19_3d}. This is used to detect collisions~\cite{Ezair2018,ahlers_19_3d}, identify compatible top surfaces~\cite{ahlers_19_3d,etienne_19_curvislicer}, and to constrain deformation gradients~\cite{etienne_19_curvislicer}. Nozzle gouging and the influence of the path orientation with respect to the slope are identified as important issues~\cite{Chakraborty_2007,allen_15_experimental,song_2016,ahlers_19_3d}, even though not directly addressed.

In comparison to the aforementioned techniques, our approach globally curves the part layers with benefits similar to Etienne et al.~\cite{etienne_19_curvislicer} at a fraction of its cost. Notably, it does not require a tetrahedrization, and does not require iterating a solver. 
In addition, most existing techniques generate toolpaths following strategies for planar layers --- in fact, the curved layers are typically obtained as the deformed toolpaths of flat layers~\cite{song_2016,ahlers_19_3d,etienne_19_curvislicer}. Instead, we explore a dedicated toolpath strategy, orienting paths according to the surface curvatures. This enables novel tradeoffs in how curved surfaces are reproduced.






\subsection{Curved printing with multi-axis machines}

The previous techniques -- as well as ours -- focus on 3-axis machines. However, to obtain more freedom and allow the nozzle to incline and follow surface slopes, another related line of research focuses on multi-axis systems. These can be for instance multi-DOF extrusion devices (e.g.~\cite{huss_23_gravitiy, srinivas_23_supportless}) or robotic arms (e.g.~\cite{lettori_24_implementation, zhang_22_s3}).

Multi-axes non-planar printing allows to improve other aspects of printed parts, such as reducing support material~\cite{dai_18_support, lau_23_partition, hong_22_open5x, liu_24_spherical}, improving structural strength~\cite{perezcastillo_23_flexural, shan_23_additive, bi_23_strength, guidetti_23_stress}, considering multiple objectives together~\cite{zhang_22_s3}. This can be combined with fiber extrusion for large improvements in part strengths~\cite{akhoundi_24_gcode, fang_24_exceptional, kipping_23_load, palmer_23_effects, zhao_23_dimension, zhang_24_improved}.


We refer the reader to the review by Tang et al. \cite{tang_24_review} for an in-depth state-of-the-art on the topic of non-planar printing, including multi-axis additive manufacturing.

\section{Method}

\begin{figure}[htbp]
    \centering
    \includegraphics[width=0.3\linewidth]{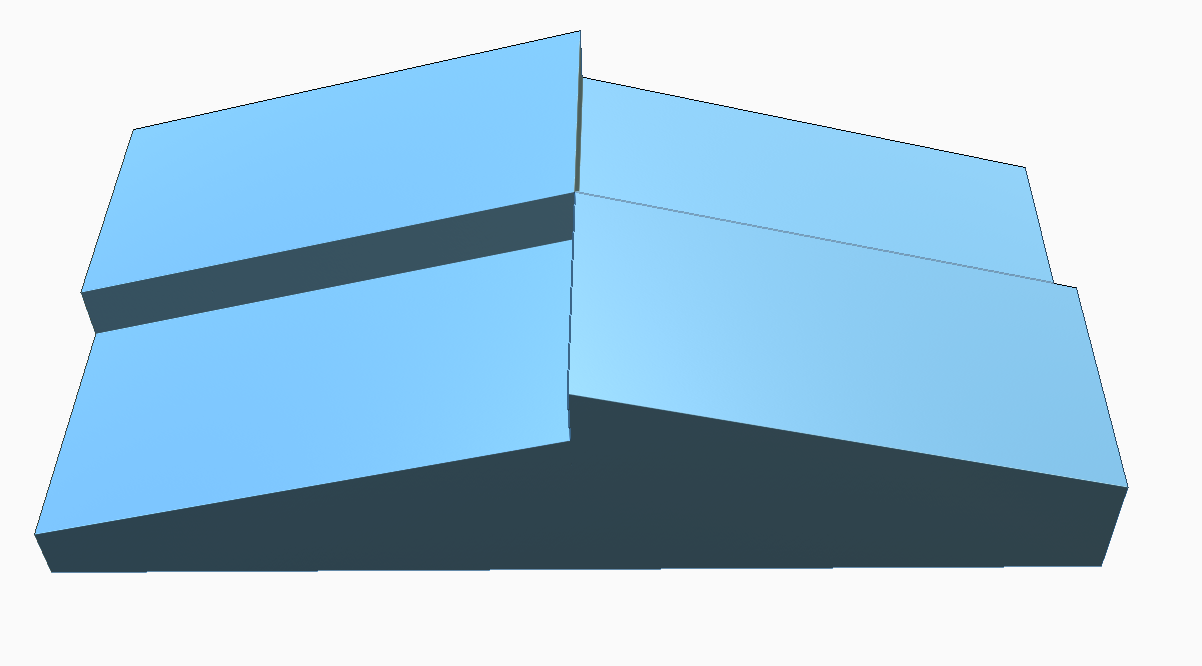}
    \includegraphics[width=0.3\linewidth]{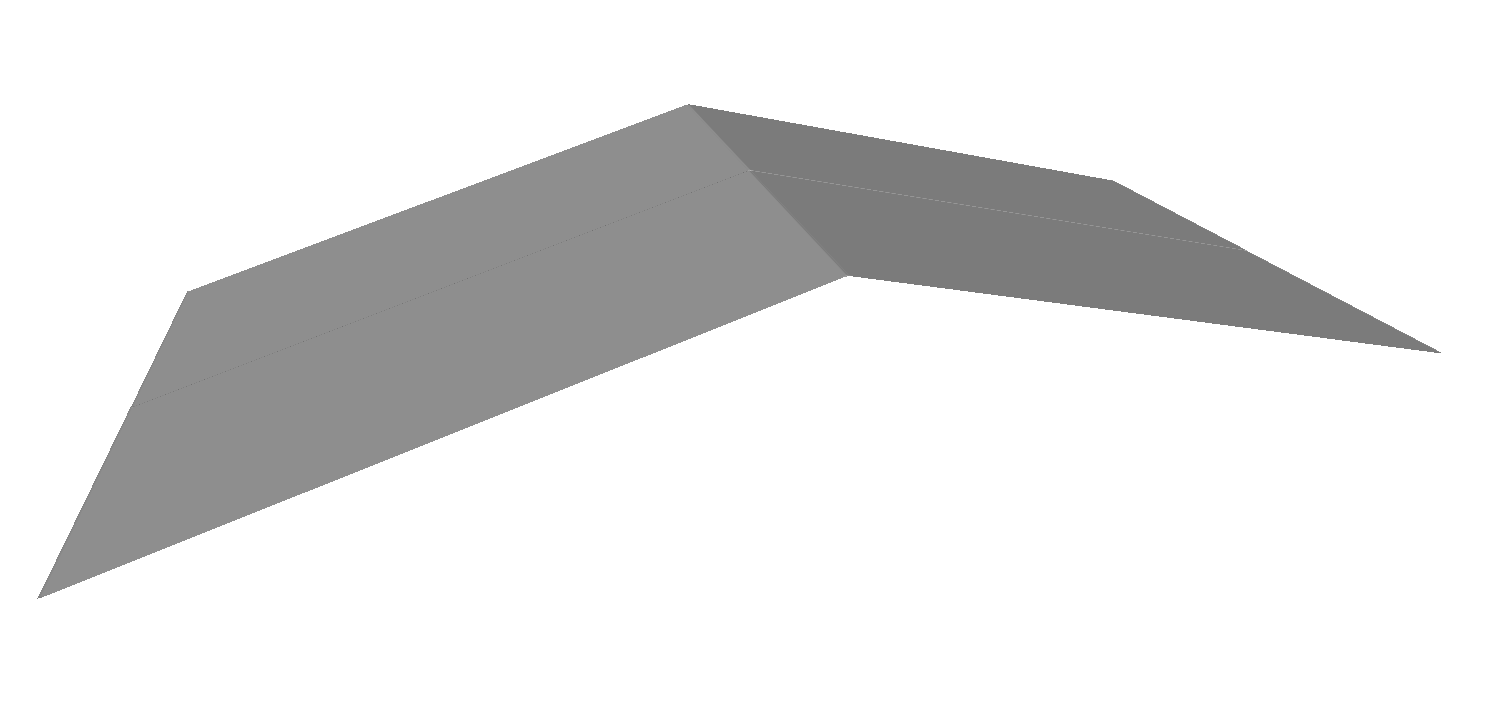}
    \includegraphics[width=0.3\linewidth]{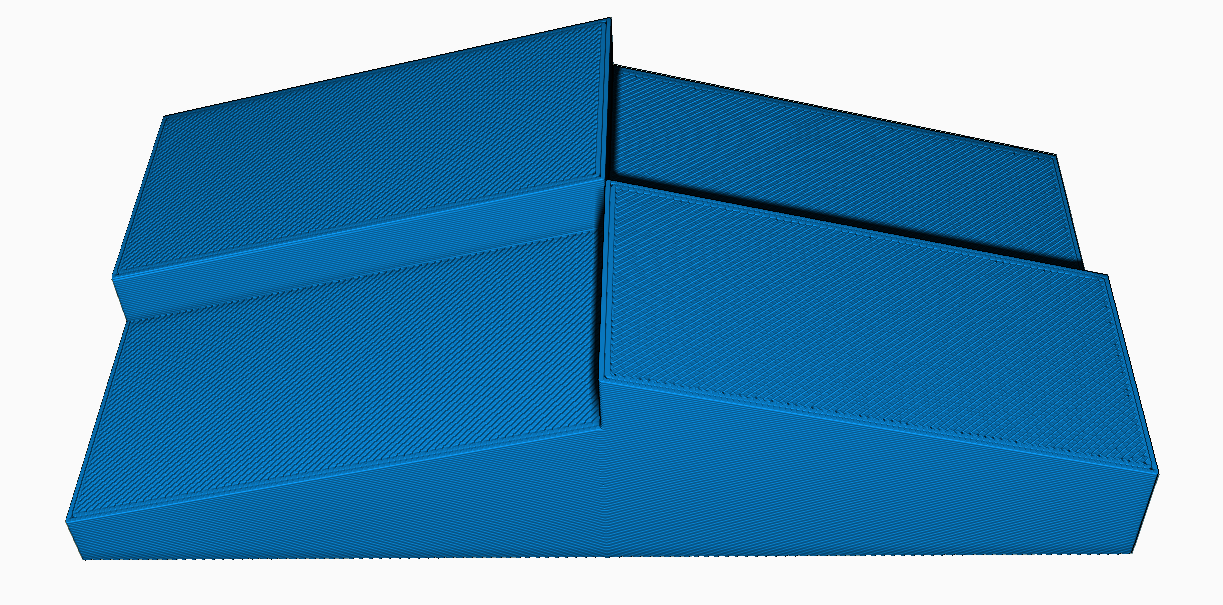}
    \caption{From left to right: A shape with sloped surfaces, the slicing surface optimized by our approach, the obtained trajectories. All surfaces are exactly reproduced (zoom for details). Note how the vertical ridges do not prevent curving the top surfaces.}
    \label{fig:simple}
\end{figure}

Our method is based on the idea of slicing the part with a single non-planar surface, see Figure~\ref{fig:teaser}. The contour of every layer is obtained by intersecting the part with this slicing surface at different heights. The slicing surface is optimized to capture the curvature of the top layers while ensuring no collisions can occur.
In particular, wherever the slicing surface slopes align with the object surface slopes no staircase defect occurs. See Figure~\ref{fig:simple} for an example.
We next describe the problem setup, our method and how it operates.

\subsection{Input geometry and output trajectories}

Our method takes as input a mesh or solid definition $\mathcal{M}$. The only requirement on $\mathcal{M}$ is to properly define a volume, such that ray-volume intersections produce well defined solid intervals.

The output is an ordered set of toolpaths. The toolpaths are polylines in $x,y,z$ (3D position). In addition, each vertex is associated with an attribute $E$ which controls the amount of material extrusion from one segment to the next. 
The toolpaths form a continuous path in $x,y,z$ with some segments performing no extrusion: they correspond to travel moves. 

We seek to optimize a slice surface $\mathcal{S}$ defined as a height field $\mathcal{S}(x,y) : \mathbb{R}^2 \rightarrow \mathbb{R}$ above the build surface. We now discuss the problem formulation in terms of constraints and objectives.

\subsection{Surface accessibility}

As discussed in Section~\ref{sec:prevwork} all surfaces having a slope\footnote{We measure slope as the angle between the surface gradient and the horizontal plane, i.e. the XY plane has a zero slope.} below a maximum slope angle $\theta_{max}$ can be accessed by the nozzle without self-collisions. This stems from the nozzle conical collision model~\cite{Ezair2018,ahlers_19_3d,etienne_19_curvislicer}.

We therefore define a \textit{valid} slicing surface as a continuous surface where everywhere the slope is below $\theta_{max}$. Such a surface has a bounded gradient~\cite{etienne_19_curvislicer} and guarantees fabricability under the conical collision model. Indeed, at any given time during fabrication the slicing surface separates what has been printed below from the layer being currently printed. As all print trajectories for a layer are along the slicing surface, and the slicing surface is itself free of self-collisions, no collisions can occur.

Note that on some printers the conical model is not sufficient as the extrusion device body may be close to the nozzle tip~\cite{ahlers_19_3d}. We ignore this for the sake of simplicity, noting that in practice the maximum slope can be lowered to account for the extruder body.

\subsection{Sloping control}

Our method is controlled by an input map $\Theta : \Omega \rightarrow \mathbb{R}$ with $\Omega \subset \mathbb{R}^2$. The map indicates in selected $x,y$ locations the ideal shape of the surfaces to be fabricated, as connected pieces of heightfields. We measure the slope of $\Omega$ at a location $x,y$ by noting $\theta(\Theta(x,y))$. This operator computes the slope through finite differences in $\Omega$. 

The map has to be given such that everywhere in $\Omega$ it respects $\theta(\Theta(x,y)) < \theta_{max}$. In the reminder, for the sake of clarity we will often refer to $\theta(\Theta(x,y))$ as simply $\theta$ where it is not ambiguous to do so.

$\Theta$ is typically a sparse map: it only defines the target surfaces in selected locations. In all our examples we obtain $\Theta$ by selecting the top surfaces of $\mathcal{M}$ -- as seen from above -- having a slope below a user-chosen target maximum slope $\theta_{target} < \theta_{max}$. However, the user could also specify the map directly or it could be optimized by another algorithm: the surfaces in $\Theta$ do not necessarily have to be a subset of those of $\mathcal{M}$.

Note that $\theta_{max}$ and $\theta_{target}$ have distinct roles. $\theta_{max}$ \textit{ensures} safe fabrication without collisions, while $\theta_{target}$ is a user preference. Indeed, due to nozzle gouging it is rarely desirable to try to curve everything below $\theta_{max}$, and instead a compromise can be preferable. We discuss these tradeoffs in Section~\ref{sec:results}.

\subsection{Slice optimization}
\label{sec:optim}

\begin{figure}[htbp]
    \centering
    \includegraphics[width=0.3\linewidth]{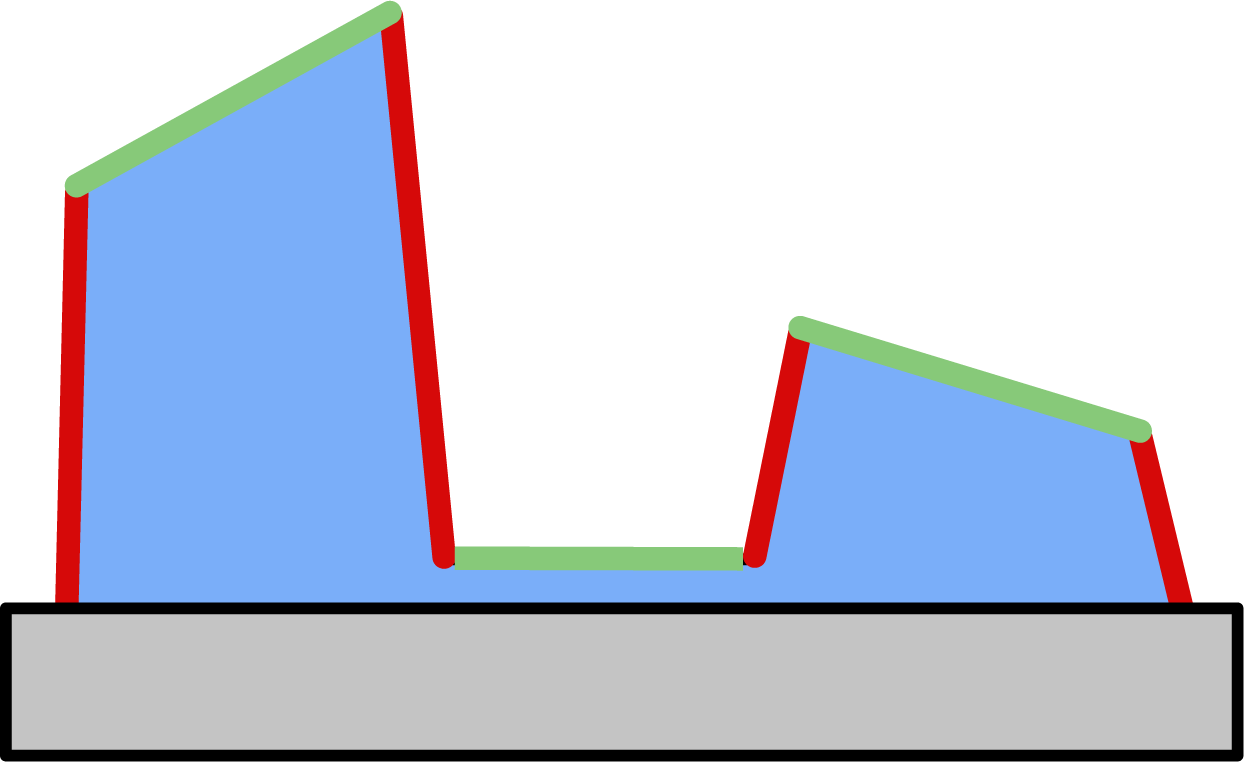}
    \includegraphics[width=0.3\linewidth]{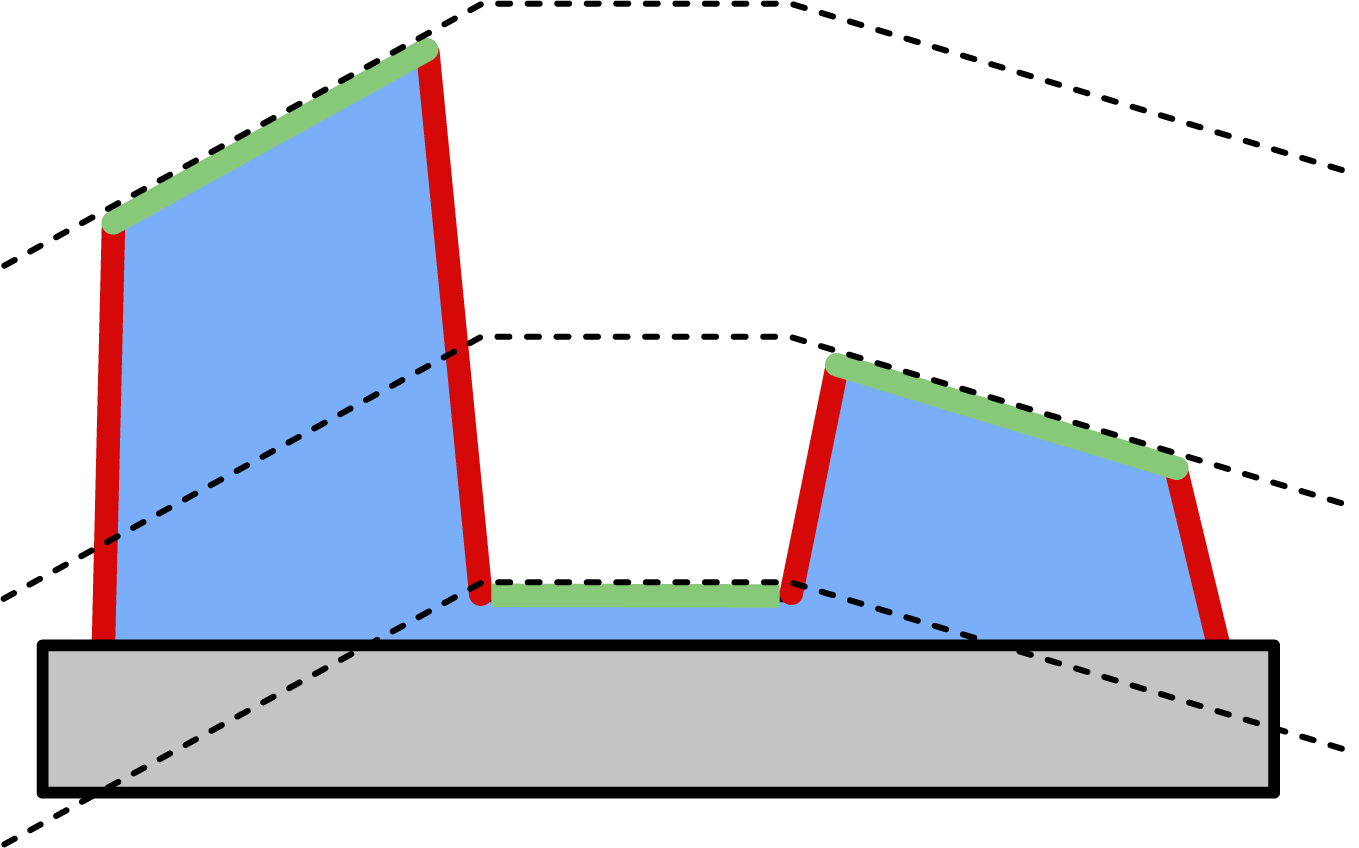}
    \caption{Left: The green surfaces are specified in $\Theta$, the red ones are free as their slopes exceed $\theta_{target}$. The relative heights of the green surfaces makes it impossible to find a slicing surface with slopes everywhere below $\theta_{max}$. Right: By allowing the green surfaces to 'float' up and down during optimization, our method computes the dashed line as a slicing surface. It exactly reproduces all surfaces in $\Theta$. }
    \label{fig:slicesrf}
\end{figure}

The objective of our algorithm is to optimize for a slicing surface used to decompose $\mathcal{M}$ in a set of layers. The primary objective is for the slicing surface to ideally reproduce the surfaces given in $\Theta$. This is achieved when, at the time of slicing, the slicing surface exactly aligns with the shape of each surface in $\Theta$ (Figure~\ref{fig:slicesrf}, right). Note that following this reasoning the absolute altitudes of the surfaces do not matter: it suffices that there exists a height where the slicing surface aligns with the target surface. This is the key observation onto which we base our optimizer.

The slicing surface should also be valid to ensure a correct fabrication. In the optimization step we consider slice validity as an objective. We later enforce validity as a post-process (Section~\ref{sec:postproc}).

We optimize $\mathcal{S}$ as a discrete height field. Therefore, the variables we optimize for are a set of displacements (or altitudes) from the printing bed, noted $h_{(i,j)}$ with $i,j$ an index in the 2D discretized version of $\mathcal{S}$.
The degrees of freedom are the altitudes were a target surface is \textit{not} specified in the control map $\Theta$. Where $\Theta$ is specified we want to enforce the slicing surface shape to follow the given surfaces.

A reasonable objective to fill-in free regions around the surfaces in $\Theta$ would be to make the slicing surface as smooth as possible in between. 
However, as illustrated in Figure~\ref{fig:slicesrf} (left) the top surfaces (in green) may be vertically separated by a large distance while being comparatively close in $x,y$. Therefore, no smooth slicing surface can be produced that would have a slope under $\theta_{max}$. This is the main reason why curving the middle surface in Figure~\ref{fig:two_towers} is challenging.

To address this issue we introduce additional variables, one per connected component in $\Theta$. Intuitively these allow each target surface to move vertically, to 'float' up and down. Let us denote the set of connected components in $\Theta$ as $\mathcal{C}$. We associate a variable $z_c$ with each $c \in \mathcal{C}$. The values of $h_{(i,j)}$ in a component $c$ are fully determined by $z_c$ as $h_{(i,j)} = z_c + \Theta(i,j) - \Theta(c_0)$, with $c_0$ a reference point arbitrarily chosen in $c$.

Every free variable neighboring a connected component is requested to match the height of its neighbor. That is, given a coordinate $i,j$ in component $c$ with a free neighbor $h_{(i,j) + \delta}$ with $\delta \in [-1,1]^2$, we add the objective that $h_{(i,j) + \delta} = z_c + \Theta(i,j) - \Theta(c_0)$.

Within the free regions, we encourage the slice surface gradient to \textit{steepen}. This is motivated by the observation in~\cite{etienne_19_curvislicer} that surfaces should be either exactly curved, or sliced at an angle as orthogonal as possible to the surface.
Given a free coordinate $i,j$ with a free neighbor $(i,j) + \delta$ with $\delta \in [-1,1]^2$ we add the objective that 

\begin{math}
\begin{array}{ll}
h_{(i,j)} - h_{(i,j)+ \delta} = H_{target} & \text{ if } \Theta(i,j) > \Theta((i,j)+\delta)\\
h_{(i,j)+ \delta} - h_{(i,j)} = H_{target} & \text{ otherwise }
\end{array}
\end{math}

\noindent Where $H_{target}$ is computed from $\theta_{target}$ given the chosen discretization step of $S$ (we typically use $50 \mu m$). We therefore request the local gradient to increase to match the selected target slope.

From this, we optimize for the free $h_{(i,j)}$ and $z_c$ to obtain a surface having the target gradient.
This leads to a least square problem similar to solving for the Poisson equation in a grid~\cite{perez_2003}, with the addition of the per-component variables. We solve it using the Eigen library~\cite{eigen}.

\begin{figure}[htb]
    \centering
    \includegraphics[width=0.3\linewidth]{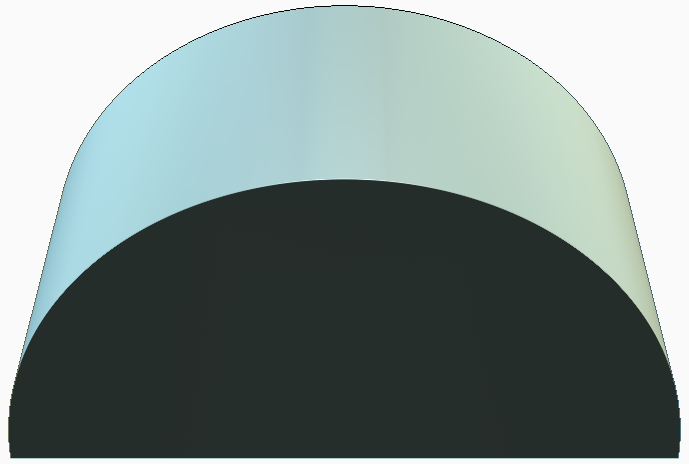}
    \includegraphics[width=0.3\linewidth]{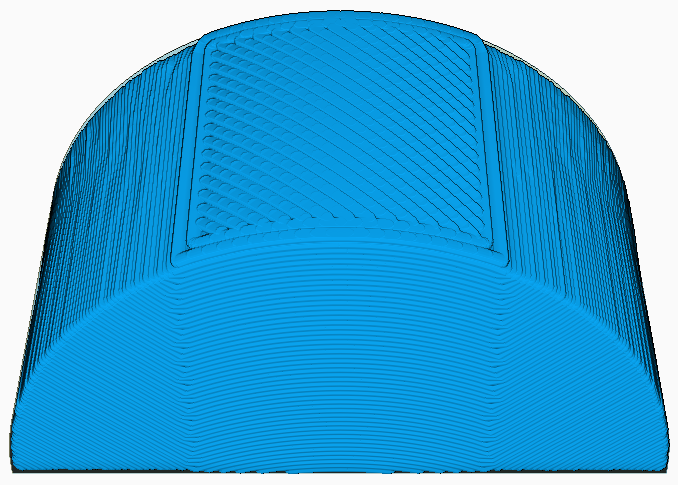}\\
    \includegraphics[width=0.3\linewidth]{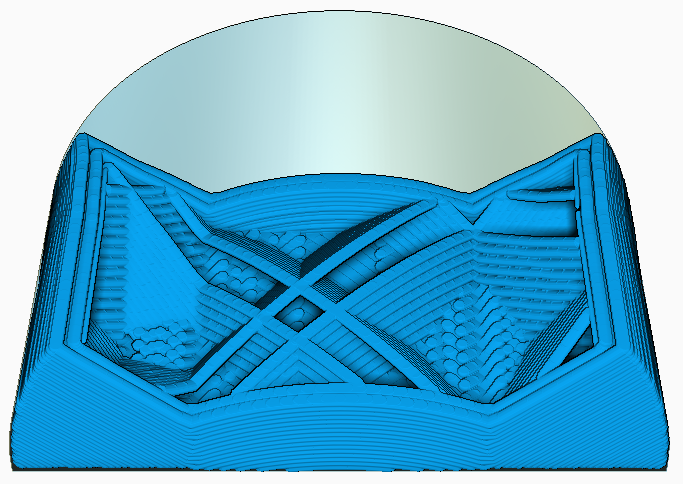}
    \includegraphics[width=0.3\linewidth]{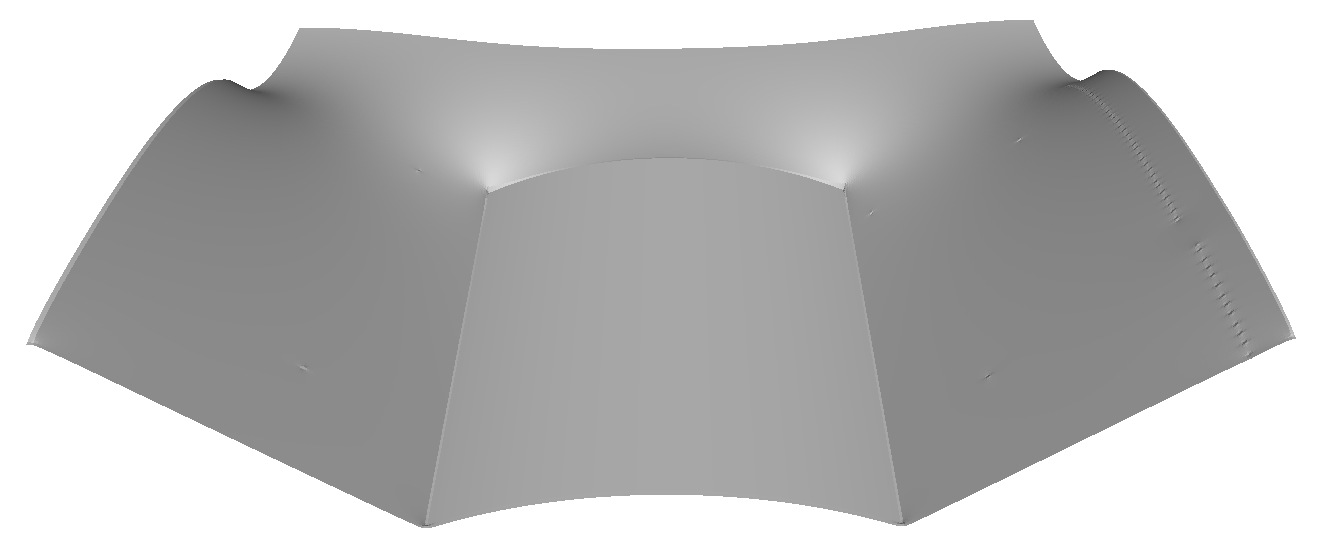}
    \caption{Top: Half a cylinder (left) and the trajectories produced by our technique (right). Note the curved top. Bottom: Showing half of the bottom layers (left) reveals the shape of the slicing surface (right). The back of the slicing surface is a smooth free region where the extruder travels when the print starts.}
    \label{fig:cyl}
\end{figure}

A first result is shown in Figure~\ref{fig:teaser} for the shape in Figure~\ref{fig:two_towers}. A half-cylinder model is shown in Figure~\ref{fig:cyl}. Here only the top of the cylinder can be curved. Thanks to gradient steepening, the slicing surface takes an orthogonal shape with respect to the rest of the surface, improving its slicing quality.


While this is often sufficient to obtain a valid slice, this formulation does not guarantee it. 
The geometry of the constraint surface and their locations in $\Theta$ can make it impossible to find such a solution.
Instead of relying on inequality constraints and a more complex solver, we propose a post-processing approach that efficiently resolve sparse constraint violations.



\subsection{Post-processing}
\label{sec:postproc}

\begin{figure}[htbp]
    \centering
    \includegraphics[width=0.19\linewidth]{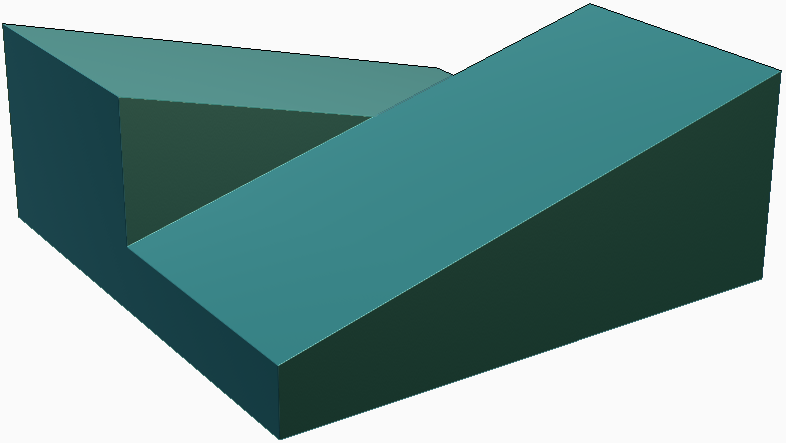}
    \includegraphics[width=0.19\linewidth]{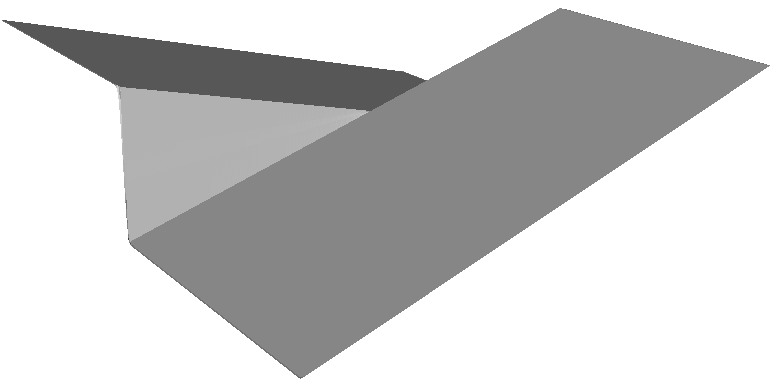}
    \includegraphics[width=0.19\linewidth]{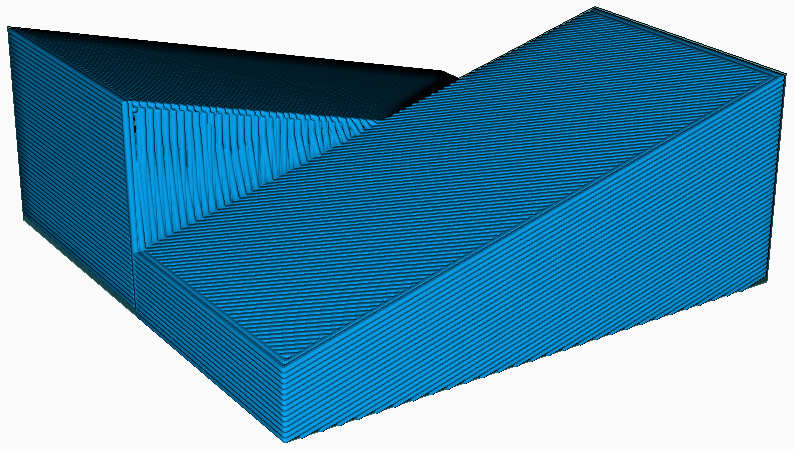}
    \includegraphics[width=0.19\linewidth]{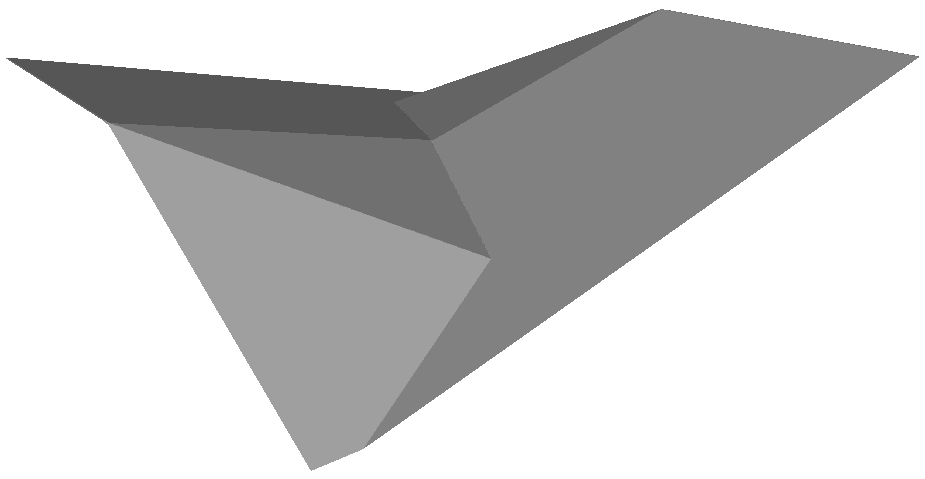}
    \includegraphics[width=0.19\linewidth]{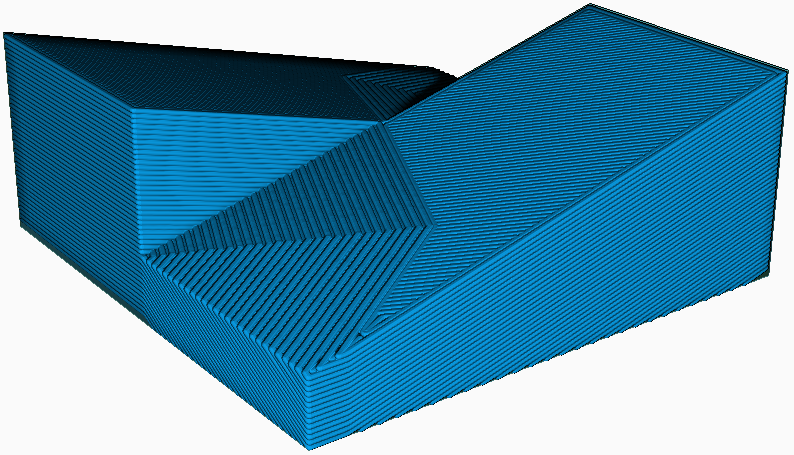}
    \caption{Shape to be sliced, Optimized slice surface (the vertical ridge violates $\theta_{max}$), The resulting GCode cannot be printed safely, The corrected slice surface (the angle everywhere complies with $\theta_{max}$), The resulting GCode (note how parts of the surfaces are now sliced to avoid collisions).}
    \label{fig:shield}
\end{figure}

The optimizer may generate a surface where the slope exceed $\theta_{max}$ in locations. This is typically rare when $\theta_{target}$ is well below $\theta_{max}$, but becomes an issue at higher values of $\theta_{target}$. An example is given in Figure~\ref{fig:shield}.

Our post-process corrects the slice surface, enforcing the constraints everywhere. The algorithm sweeps the surface from top to bottom, visiting all altitudes in $\mathcal{S}$ in descending order, modifying the surface progressively. This is inspired by the algorithm of Hornus et al.~\cite{hornus_2018} for growing cavities of prescribed angles inside parts.

Let us consider a given height $h$ and a current surface $S_{h+\Delta}$, where $h+\Delta$ was the previous height. Without loss of generality, let us assume that the height $h$ exists only at a location $p$ (we simply iterate over all). We obtain $S_h$ from $S_{h+\Delta}$ by updating the locations around $p$ as $S_h(p) = max_{\delta \in [-1,1]^2}( S_{h+\Delta}(p + \delta) - H_{max}, S_{h+\Delta}(p) )$, where $H_{max}$ is the maximum allowed height difference between neighbors in the map (Section~\ref{sec:optim}). This propagates an angular constraint from top to bottom, filling areas that would have exceeded the conical constraint.

\subsection{Slicing and toolpath orientation}
\label{sec:slicing_toolpaths}

Our slicer is based on a ray-rep representation. Rays are thrown from above through the part $\mathcal{M}$, resulting in a list of solid and empty intervals along each ray. When slicing normally, contours are extracted by considering which rays are solid at the slice altitude. This makes it trivial to change the shape slice from being planar to being a height field, simply displacing the intervals along the rays according to $\mathcal{S}$.

Once obtained, the slice contour is treated as if flat. We fill the 2D outline with toolpaths, which are later deformed back as in prior works~\cite{ahlers_19_3d,etienne_19_curvislicer}.

However, different to prior works we generate toolpaths that cover the surface based on its curvature. We build upon the work of Chermain et al.~\cite{Chermain2023Orientable} who generate orientable trajectories for anisotropic appearance fabrication. In our case, we define the direction field from the curvature in $\Theta$. Our implementation reproduces the phasor optimizer and trajectory extraction described in this work.

We constrain the direction of the paths everywhere along the boundary of $\Theta$, and on a coarse grid inside (every $3$ mm) to leave some freedom to the orientation optimizer in between. Everywhere else the direction is free. We also leave the orientation free in flat regions of $\Theta$ (angle near zero). We constrain the phase in at least one point, and change the phase every layer by an angle of $\frac{\pi}{4}$ to create a staggered pattern.

\begin{figure}
    \centering
    \includegraphics[width=0.5\linewidth]{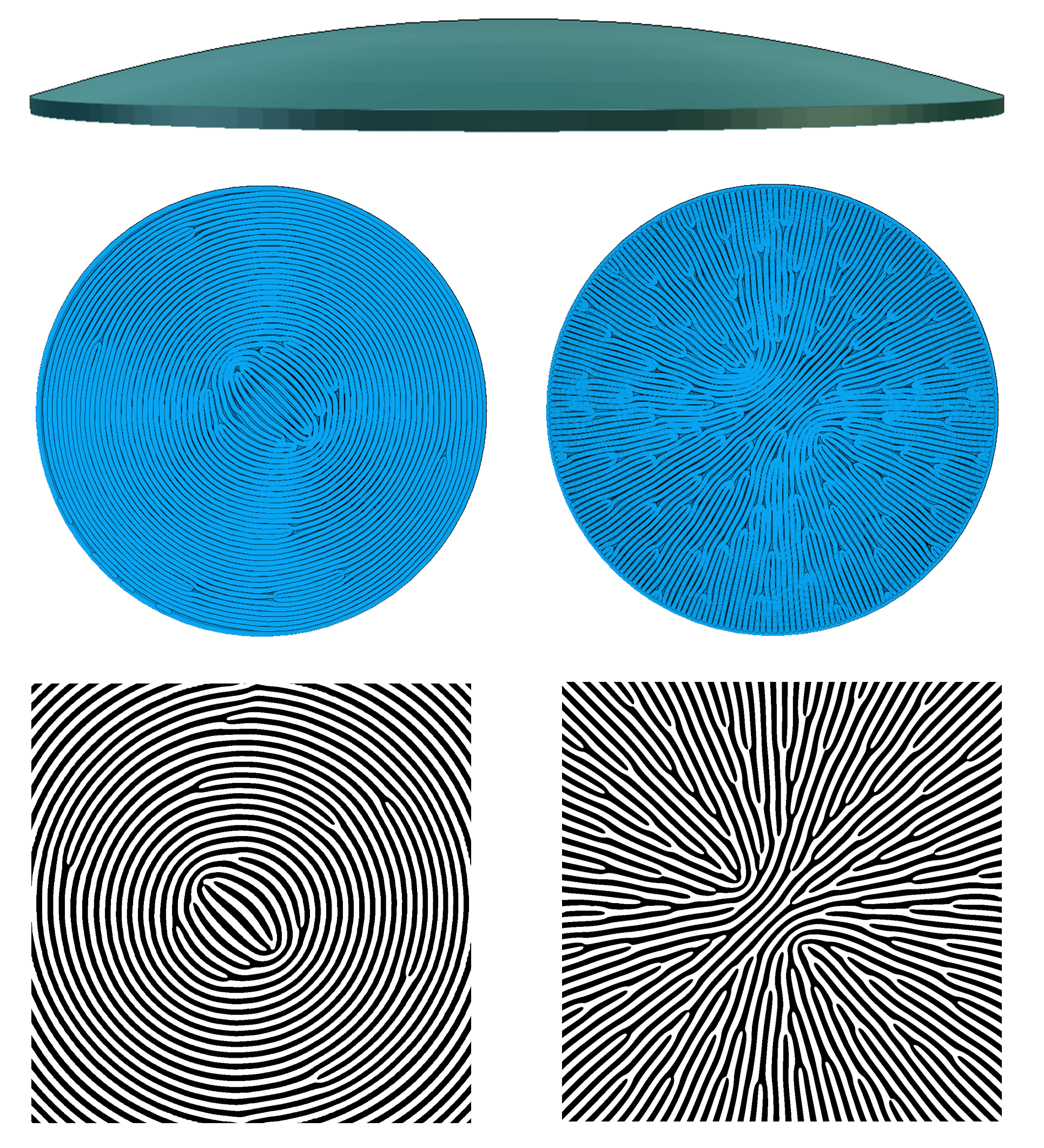}
    \caption{Top: A curved dome. Middle: The produced trajectories. Bottom: The optimized phasor fields~\cite{Chermain2023Orientable}.}
    \label{fig:phasor}
\end{figure}

This allows to produce fully filled curved surface where the orientation of the paths is precisely controlled. We explore mainly two choices of orientation with respect to the curvature.
A first choice is to orient the paths along the maximum principal curvature. This avoids create a staircase across path widths by ensuring the paths move up and down along the main slope. This is the preferred choice when seeking to fully curve a given surface.
A second choice is to orient the paths along the minimum principal curvature. Here, the paths tend to be across the main slope. We find this a good choice when the curved surfaces have a small slope and transition to parts that are printed in a non-planar fashion. 
Figure~\ref{fig:phasor} illustrates the approach on a simple case. Several examples of both choices are given in Section~\ref{sec:results}, where the method is used on all our prints.

\subsection{Filtering}
\label{sec:filter}

\begin{figure}[htbp]
    \centering
    \includegraphics[width=0.2\linewidth]{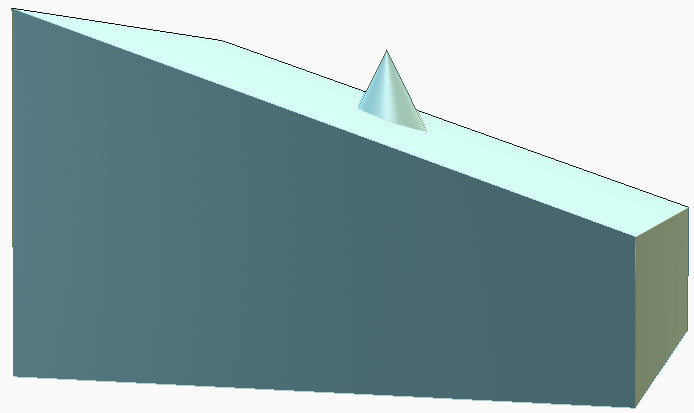}
    \includegraphics[width=0.2\linewidth]{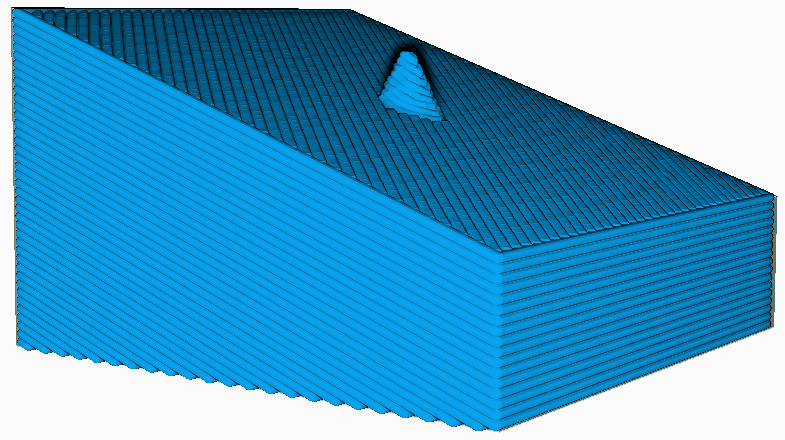}
    \includegraphics[width=0.2\linewidth]{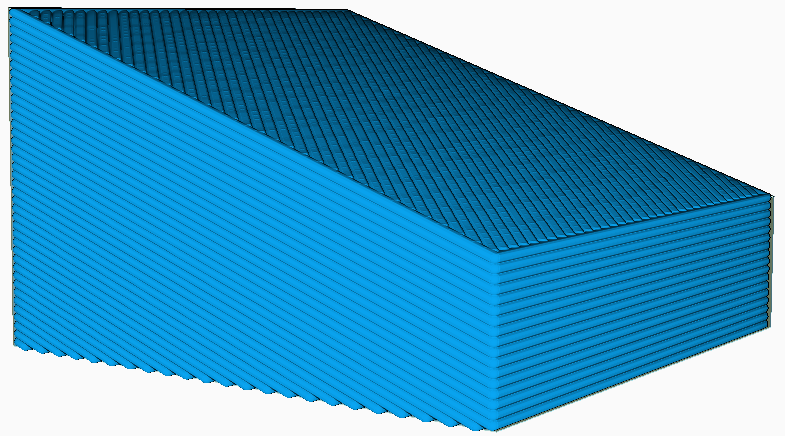}
    \caption{Illustration of filtering on a test case. This model (left) contains a spike preventing a fully curved surface (middle). This is akin to noise in scanned or natural shapes. Our filter allows to remove these tiny features, preserving the smoothness of the original surface (right).}
    \label{fig:filter}
\end{figure}

Some shapes have noisy surfaces and tiny triangulation defects, for instance scanned 3D models or tesselation artifacts. These can produce isolated island leading to spurious contours within an otherwise smooth surface, see Figure~\ref{fig:filter}. This is especially problematic when such defects have a size near the deposition width and could not be properly captured in any case.

We filter these 'holes' by performing a morphological closure on the connected components of $\Theta$. Each closed area is tagged. During the optimization of the slicing surface we disable gradient steepening in these areas, instead optimizing for smoothness. 
During slicing, in each closed area we snap the top of the shape to perfectly align with the slicing surface, canceling the geometry these features would otherwise produce.

\begin{figure*}[htbp]
    \centering
    \includegraphics[width=0.2\linewidth]{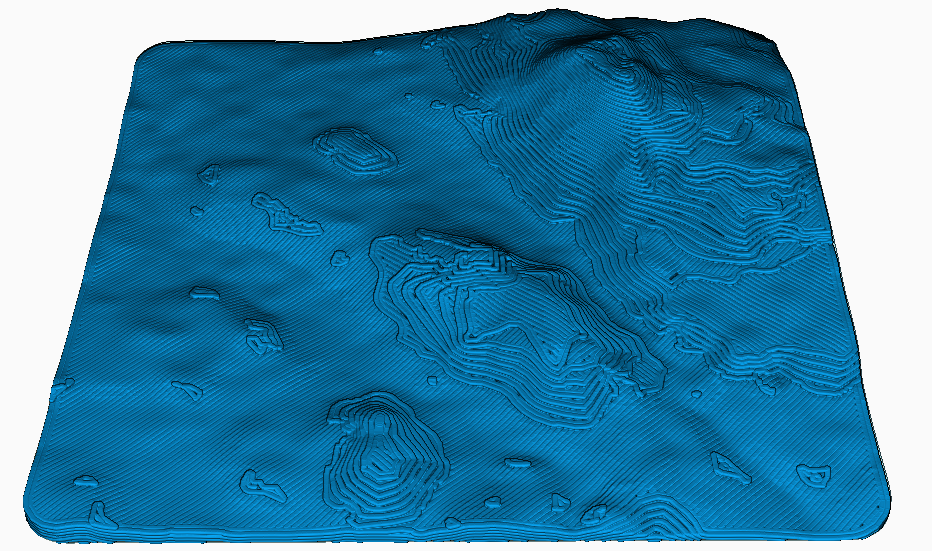}
    \includegraphics[width=0.2\linewidth]{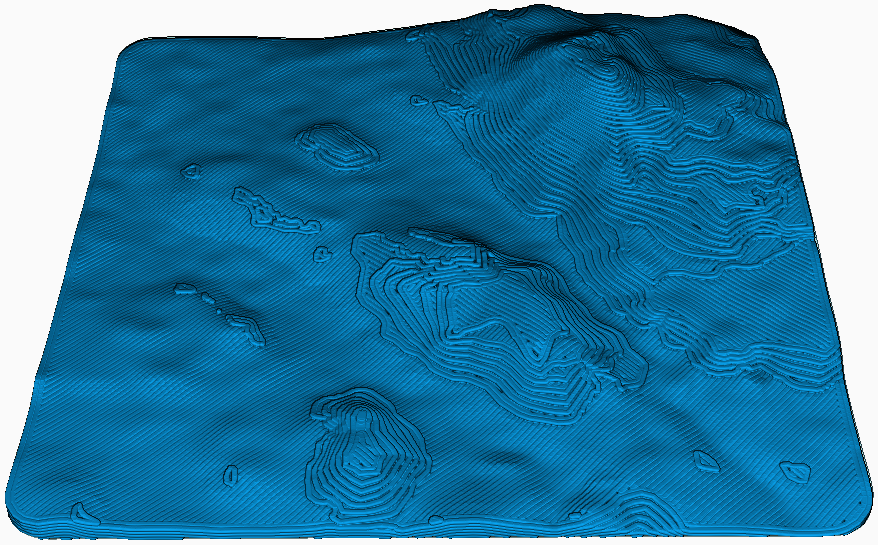}
    \includegraphics[width=0.2\linewidth]{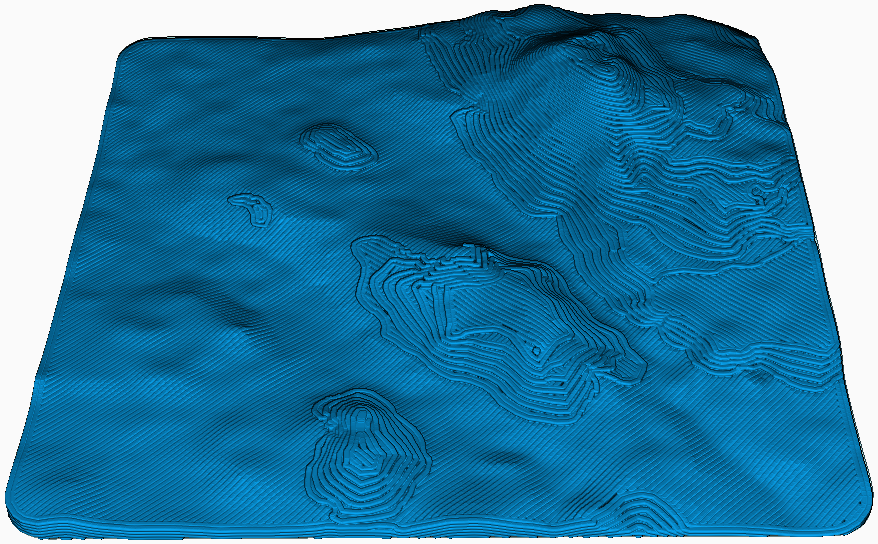}
    \includegraphics[width=0.2\linewidth]{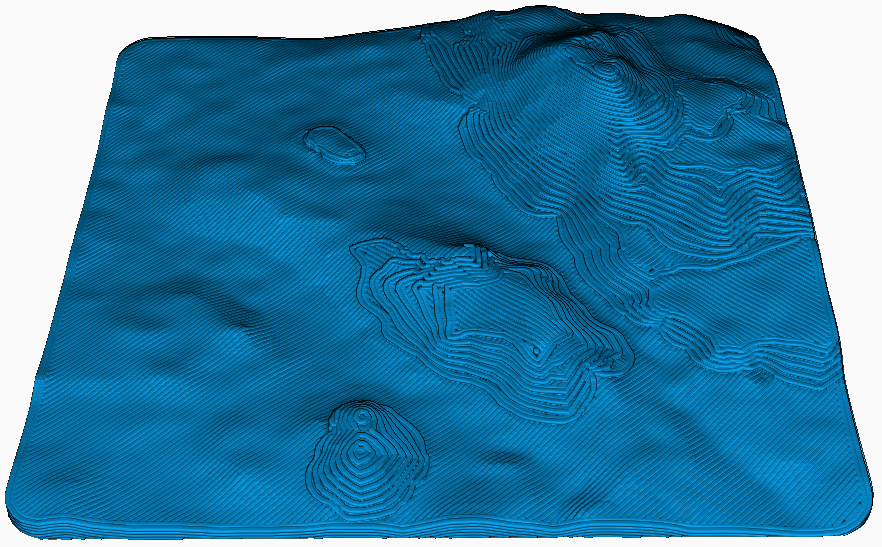}
\caption{Illustration of filtering on a real case. This terrain model (Figure~\ref{fig:map}) has a slight noise on its surface producing tiny features and spurious contours atop the main curved region. From left to right the filter radius is respectively $\rho=0$, $\rho=1$, $\rho=2$, $\rho=3$ mm. Our filter cleans up most of these issues. It is however a compromise as it modifies the geometry, for instance an overly aggressive filter (rightmost) removes the dent along the front edge from the geometry.}
\label{fig:filter_terrain}
\end{figure*}

The filtering process is driven by a single parameter, the radius $\rho$ of the closure. While optional, we find the filter important in practice on organic or natural shapes, as discussed in Figure~\ref{fig:filter_terrain}. On mechanical or regular parts using a small filter corresponding to the deposition width ($\rho = 0.2$mm) helps remove tiny regions due to tessellation and discretization noise. We precise the choice of this value in all results where it is non zero. Note that it would be straightforward to create a GUI where the user can selectively apply the filter.

\begin{figure}[htbp]
    \centering
    \includegraphics[width=0.25\linewidth]{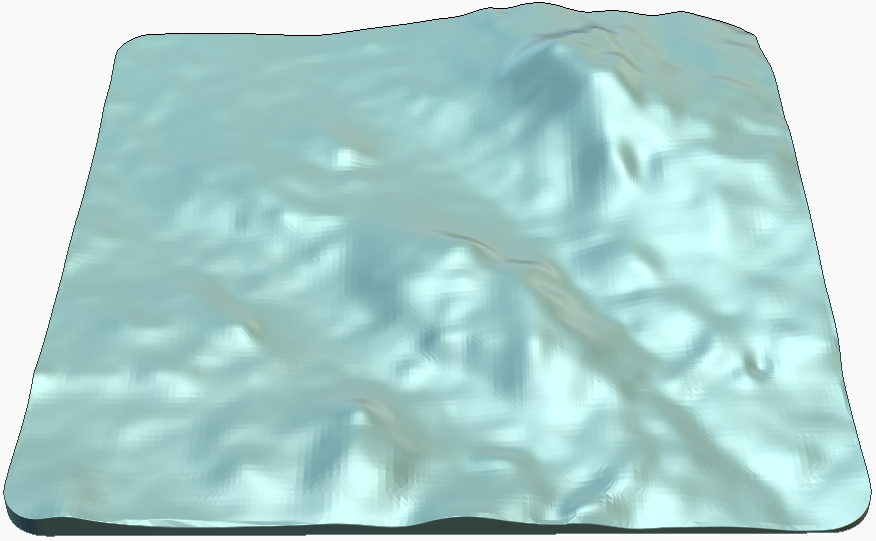}
    \includegraphics[width=0.25\linewidth]{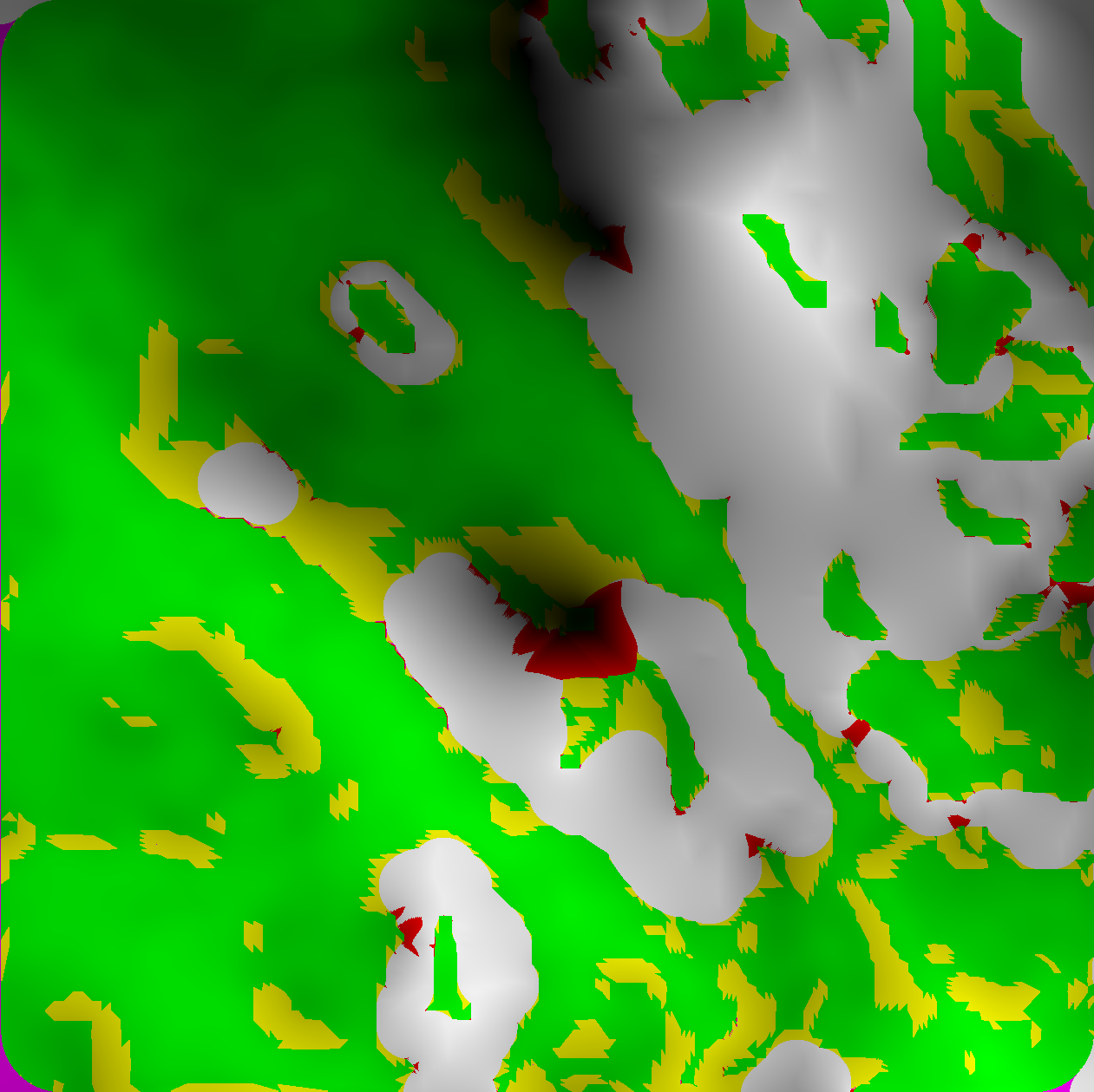}
    \includegraphics[width=0.5\linewidth]{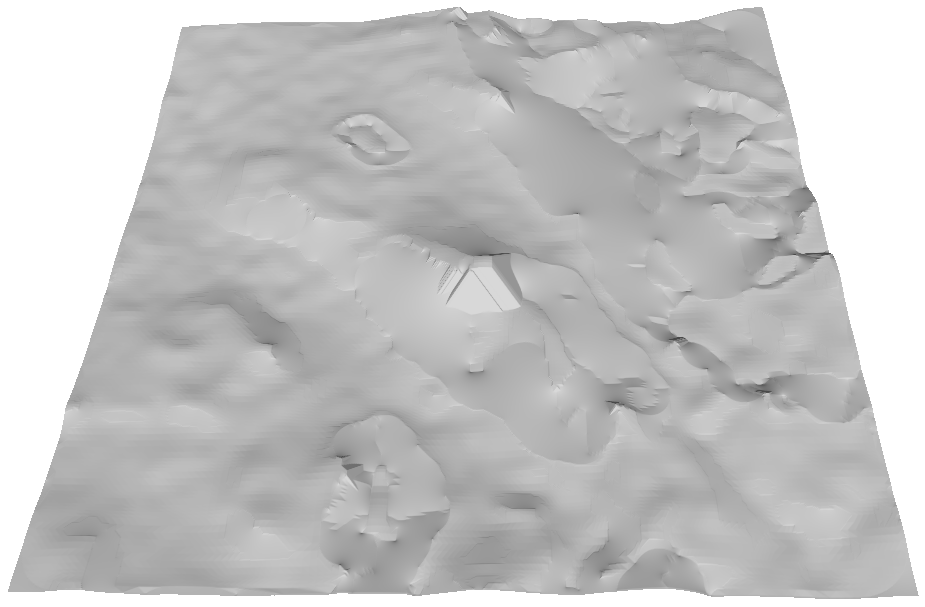}
    \includegraphics[width=0.5\linewidth]{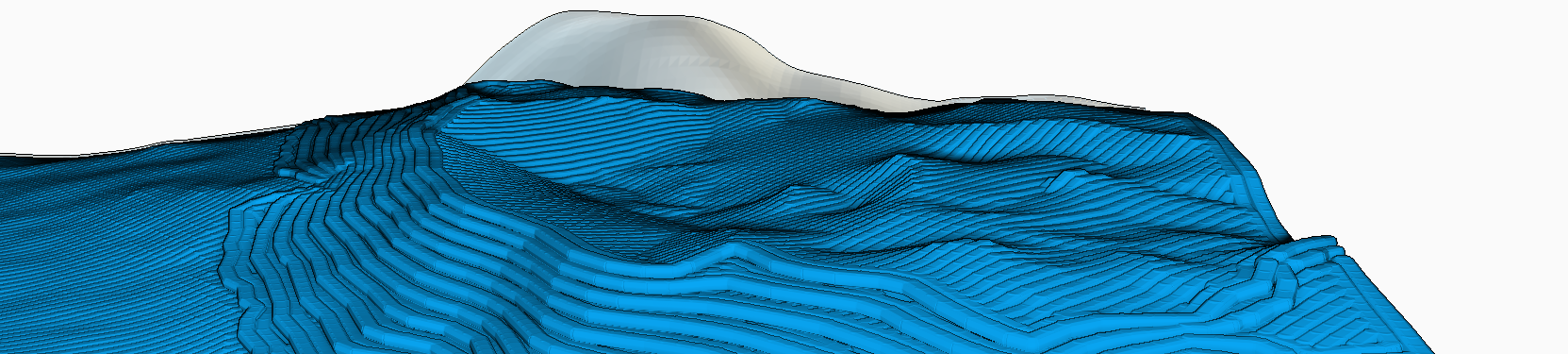}
\caption{At the top, the terrain model (top left) and a map of the various ingredients being used in various locations (top right). In green the target surface of $\Theta$, in yellow the areas impact by the filter ($\rho=2$mm), in red the regions modified by the post-process (Section~\ref{sec:postproc}). In the middle, the final slicing surface. The produced trajectories are visible in Figure~\ref{fig:filter_terrain}, third from the left. At the bottom, a close up of the mountains without the top layers, revealing the effect of gradient steepening: the interior is concave.}
\label{fig:map}
\end{figure}

The map in Figure~\ref{fig:map} shows the various ingredients used on the landscape model, next to the model itself. 

\section{Results}
\label{sec:results}

In this section we present further 3D printed results, discuss different tradeoffs, and compare to existing methods. 

\subsection{Printed results}

We use a delta printer with a nozzle offering extra clearance. Both are standard and commercially available. We print in PLA material using 1.75mm filament. The nozzle extrudes at a typical 0.4mm diameter.

\begin{figure*}
\includegraphics[width=\linewidth]{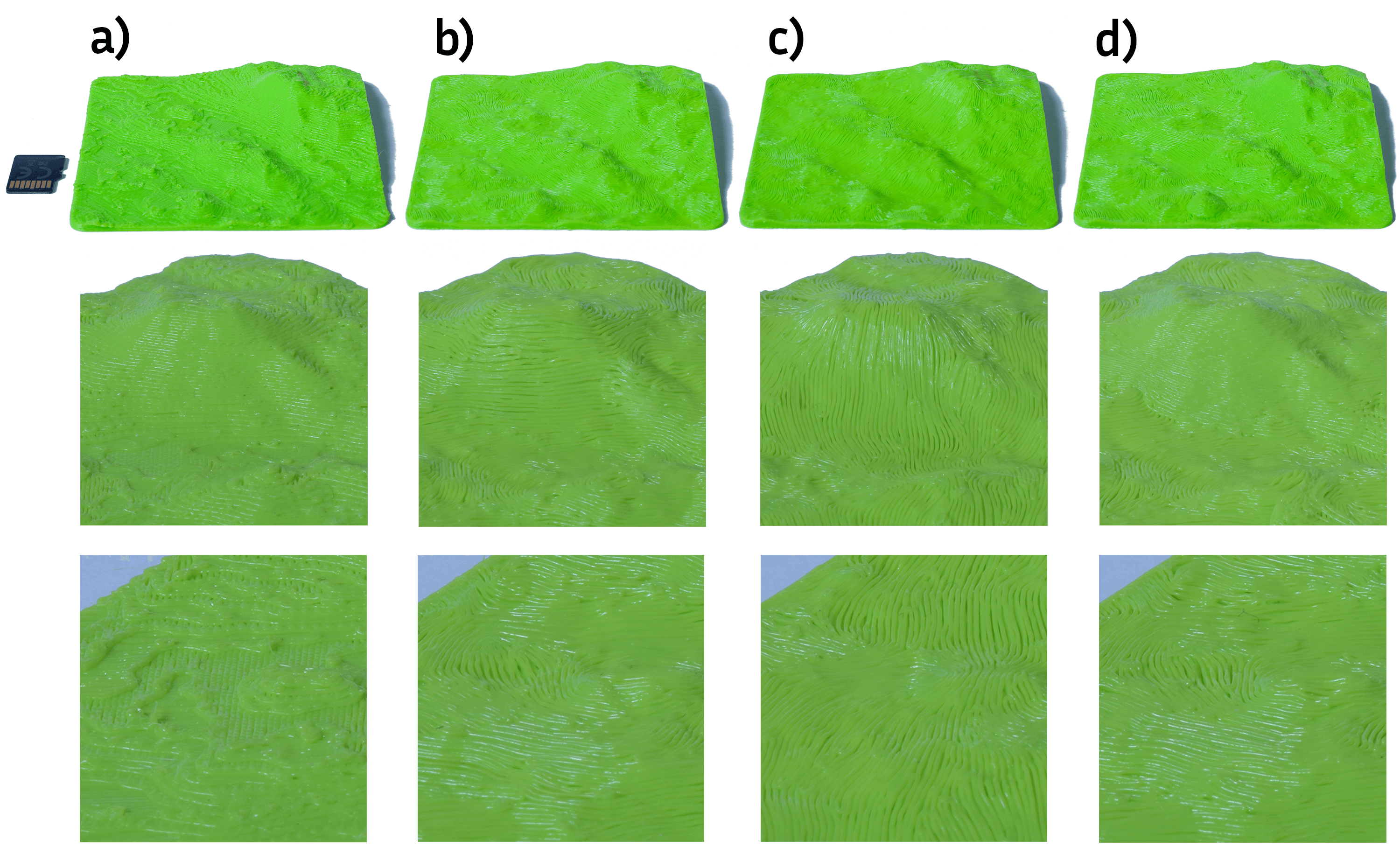}
\caption{\textit{Billinge terrain model}. a) planar print, b) ours fully curved with orientation along minimum principal curvature, c) ours fully curved with orientation along maximum principal curvature, d) ours with a lower target angle to obtain a compromise and reduce gouging. The first row of closeups focuses on the mountains. The second row focuses on a gently sloped area. (Zoom in to reveal details).
}
\label{fig:billinge}
\end{figure*}

Figure~\ref{fig:billinge} takes a closer look at different tradeoffs for printing a terrain model. This reveals how the orientation with respect to the curvature can help producing a better surface finish in most parts -- in particular where the surface is gently slopped. However, as revealed by the closeup the tall mountain suffers most from gouging where it is steepest, while standard layers captures it better. Our final result (rightmost) shows how our method can find a compromise. Here orienting the paths along the minimal curvature leads to a less noticeable transition at the base of the mountain. Note that even though the mountain is sliced, it is not a flat slicing and does benefit from gradient steepening, see also Figure~\ref{fig:map}, bottom.

\begin{figure}
\includegraphics[width=\linewidth]{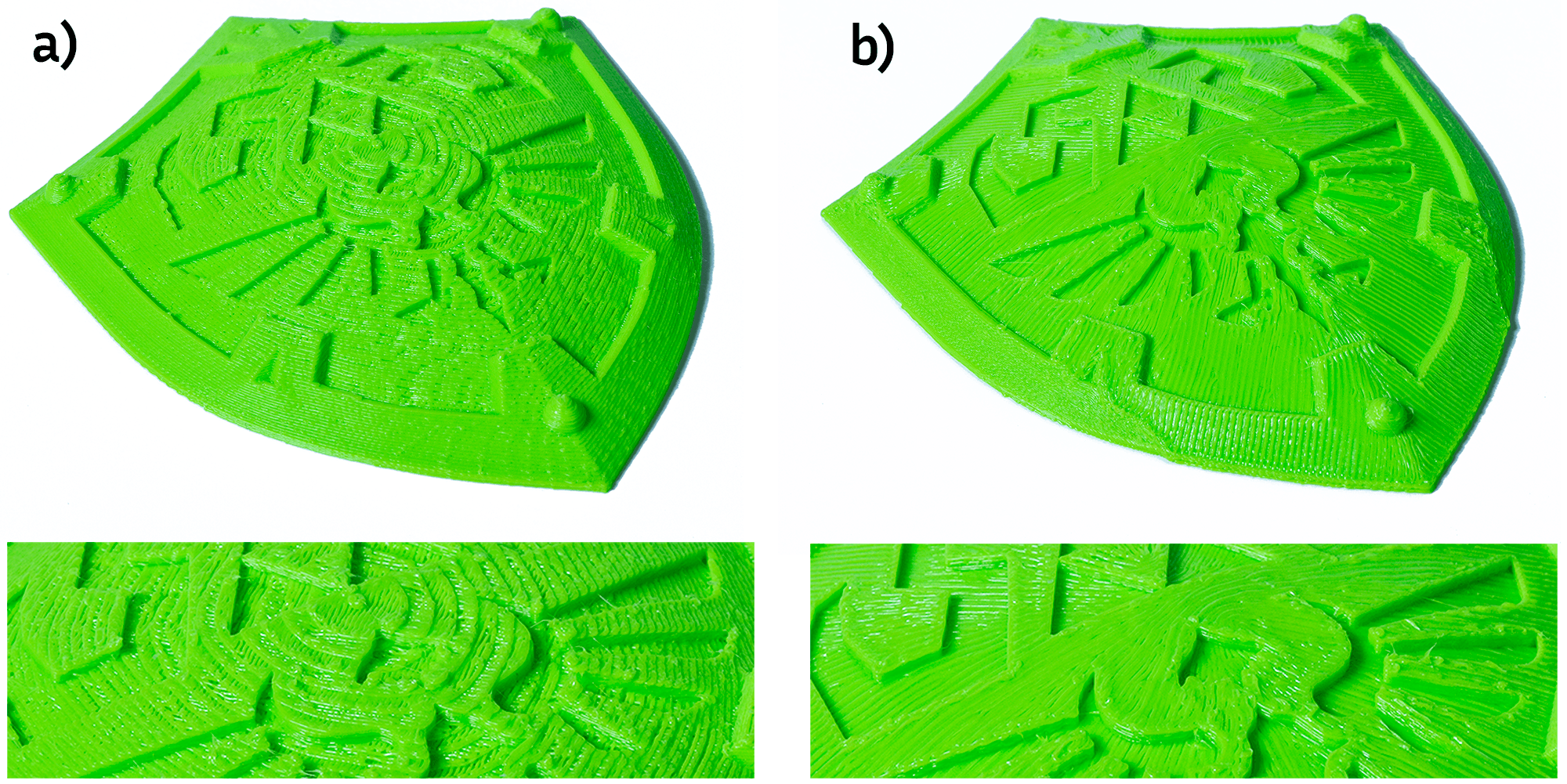}
\caption{
\textit{Zelda shield model}. a) Planar print. b) Ours with a compromise between curving and steepening ($\theta_{target} = 27\degree$, $\rho = 0.5$mm). The orientation follows the maximum principal curvature. (Zoom to reveal details).
}
\label{fig:zelda}
\end{figure}

Figure~\ref{fig:zelda} shows a Zelda themed badge, comparing the standard print to ours. Notice the significant increase in surface quality at the top, and how the trajectories naturally flow along the slope (orientation along maximum curvature).

\begin{figure}
\includegraphics[width=\linewidth]{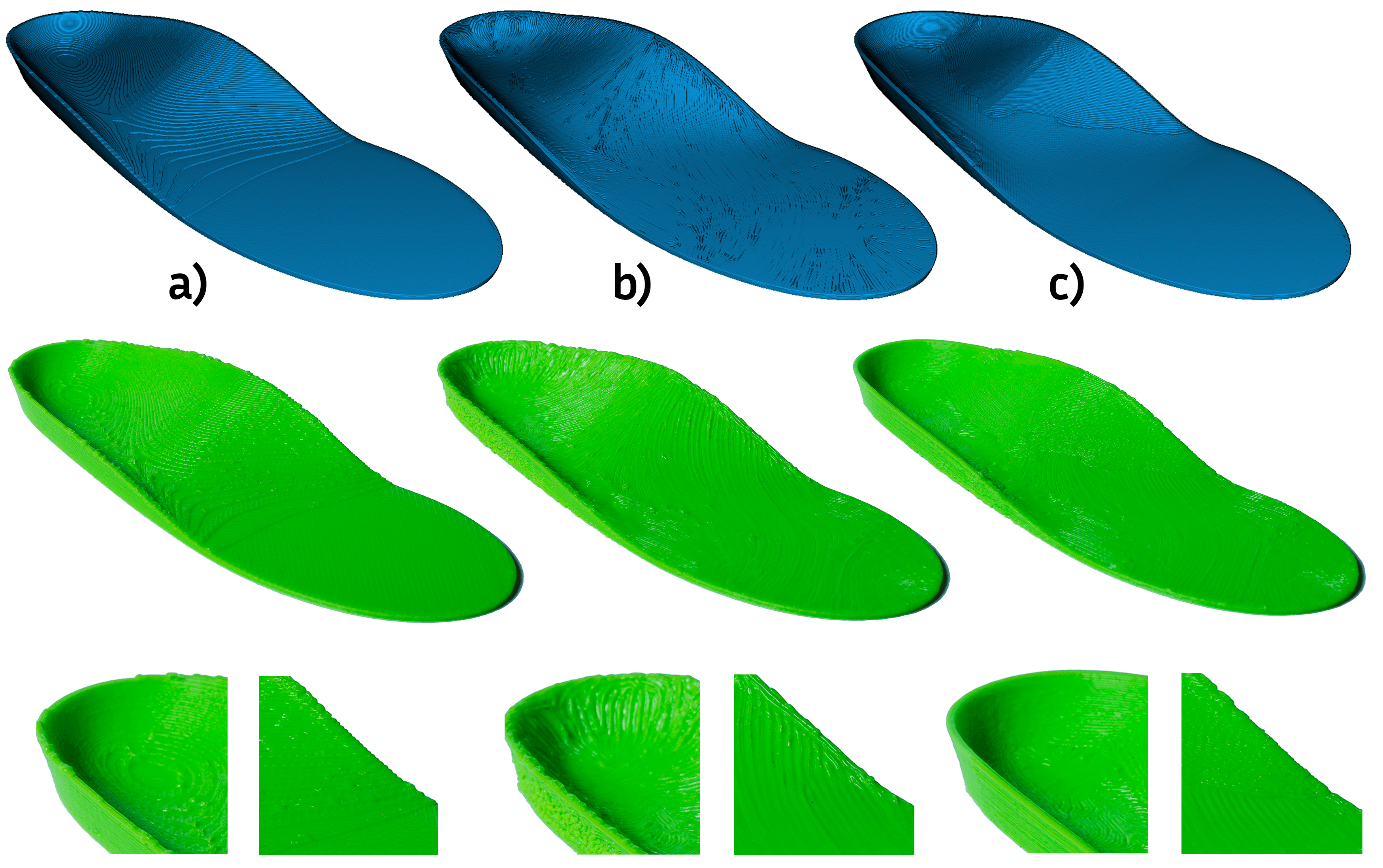}
\caption{
\textit{Insole model}. a) Planar print. b) Ours fully curved, note the gouging towards the heel. c) Ours with a compromise between curving and slicing ($\theta_{target} = 27\degree$, $\rho = 0.5$mm). On both b) and c) the orientation follows the maximum principal curvature. (Zoom in for details).
}
\label{fig:insole}
\end{figure}

Figure~\ref{fig:insole} is another example of striking a compromise between sliced and curved surfaces. While both non-planar results significantly improve the slightly sloped parts, the fully curved result suffers significant gouging near the heel. The final result captures all parts of the insole properly -- it also feels smoother to the touch in all locations.

\begin{figure}
\includegraphics[width=0.49\linewidth]{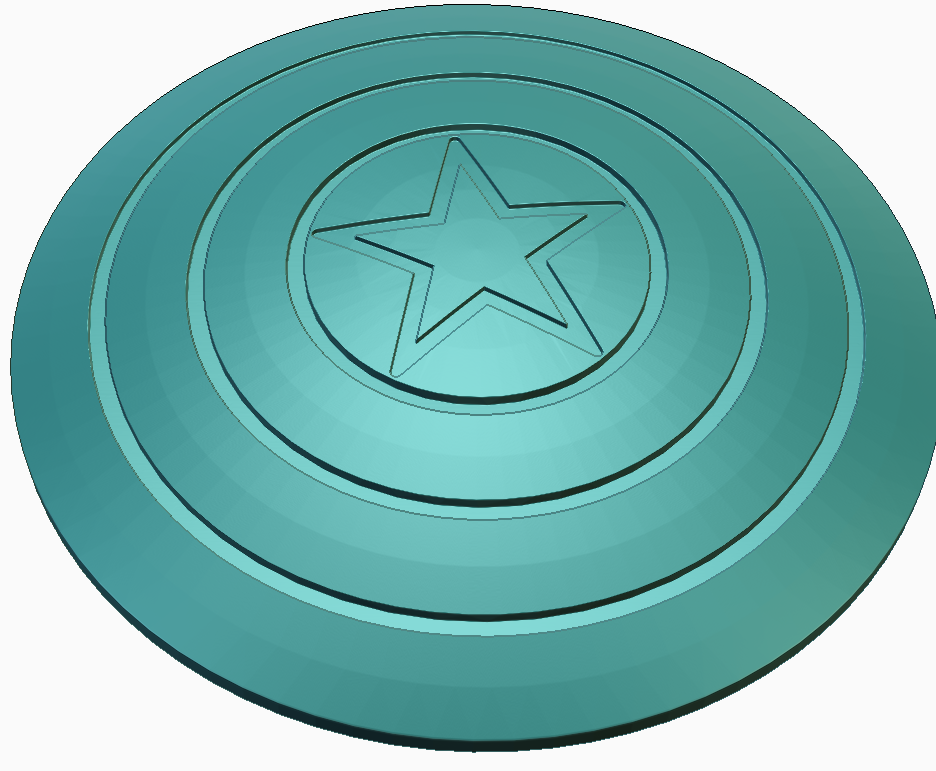}
\includegraphics[width=0.49\linewidth]{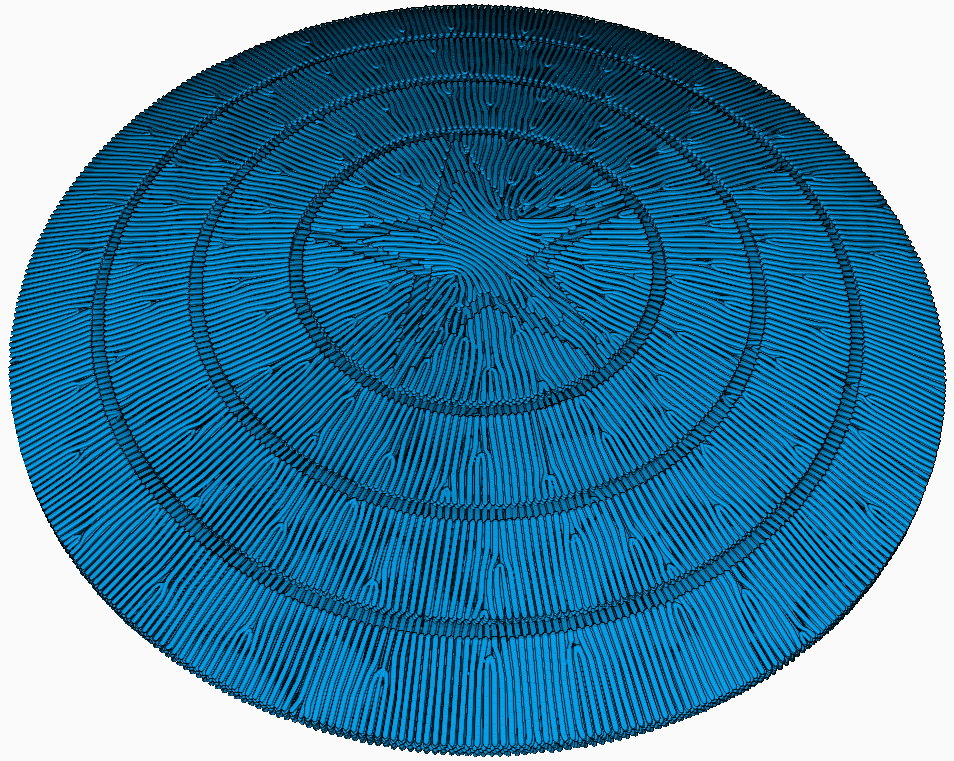}
\caption{
\textit{The Captain's shield}. A round shield model and the paths generated along the maximum curvature.
}
\label{fig:marvel}
\end{figure}

\begin{figure}
\centering
\includegraphics[width=0.5\linewidth]{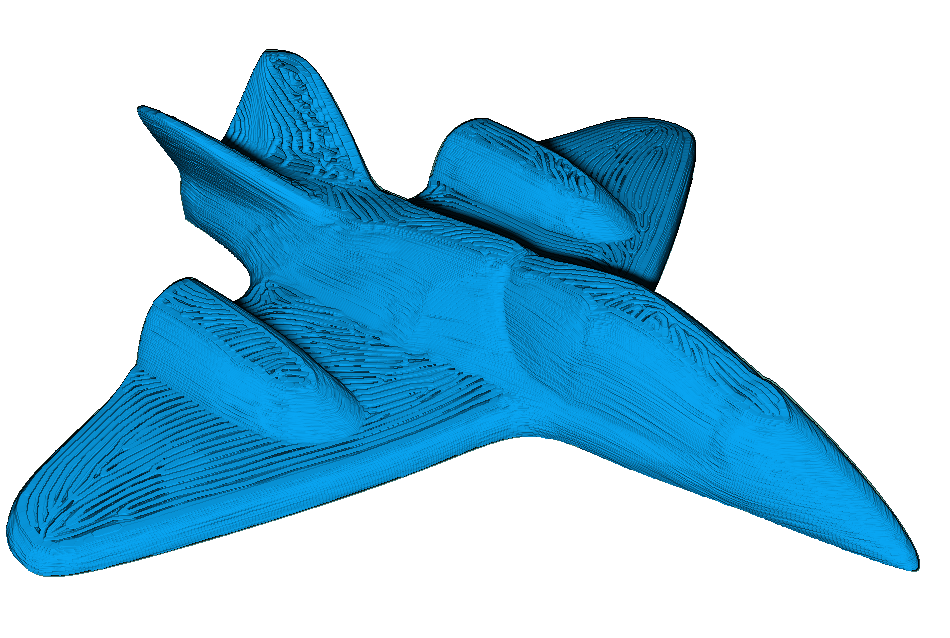}
\caption{
\textit{Jet}. The trajectories of the jet model shown in Figure~\ref{fig:teaser}, rightmost.
}
\label{fig:jet}
\end{figure}

\begin{figure}
\centering
\includegraphics[width=0.5\linewidth]{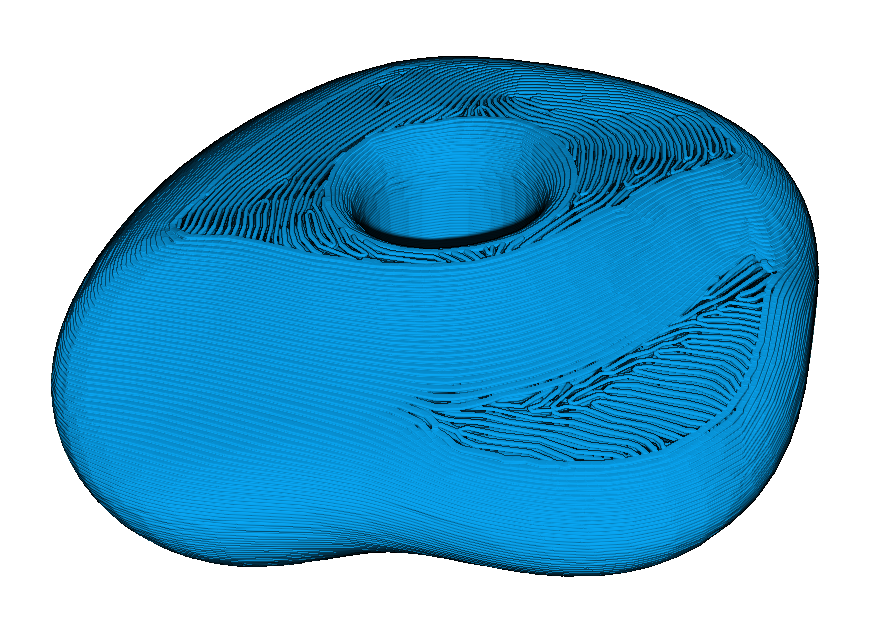}
\caption{
\textit{Climbing knob}. The trajectories of a climbing know model, note how two top surfaces are curved and separated by sliced surfaces.
}
\label{fig:knob}
\end{figure}

Figure~\ref{fig:marvel} shows trajectories optimized along the maximum principal curvature of a shield model, where all surfaces will print curved including inside the circular grooves.
Figure~\ref{fig:jet} is showing a preview of the trajectories of the jet model presented in the teaser (Figure~\ref{fig:teaser}).
Figure~\ref{fig:knob} is a climbing knob model with curved tops.



\subsection{Comparisons}

We focus our comparison on CurviSlicer~\cite{etienne_19_curvislicer} and the method of Ahlers et al.~\cite{ahlers_19_3d}. In both cases we use the publicly available implementations on github, from the original authors. We match the parameters of all methods regarding the sloping constraints and slicing parameters.

We start by comparing the computation time between Curvislicer~\cite{etienne_19_curvislicer} and our method, on several test parts found either in our submission or in~\cite{etienne_19_curvislicer}. The results are summarized in Table \ref{tab:times}. 
A computer equipped with an Intel(R) Core i7-11800H CPU (3.80 GHz) and 16 GB of RAM has been used. 
The GPU is an integrated Intel UHD Graphics.

As can be seen, our slicing optimizer outperforms Curvislicer significantly. In some instances CurviSlicer did not return a result (e.g. Jet, Insole, Billing).
While our approach is generally much faster, there is however a significant cost to using the path orientation optimizer (\textit{optizor}), and in some cases our total cost is larger due to the path orientation strategy. However, we are currently blindly applying the orientation optimizer to each and every layer, while it is most useful only on the top layers. In addition the cost varies significantly depending on the GPU, and these tests were performed on an low-power laptop GPU. 
Thus, we are confident this can be sped up significantly.

\begin{table*}[h]
\centering
\scriptsize
\tabulinesep=0.5mm
\begin{tabu}{|c|*{10}{c|}}
\hline
 & \multicolumn{4}{c|}{Curvislicer} & \multicolumn{3}{c|}{\makecell{QuickCurve \\ Standard toolpaths}} & \multicolumn{3}{c|}{\makecell{QuickCurve \\ Optizor toolpaths}} \\
\hline
 & \makecell{Tetrahed-\\-ralization} [s] & \makecell{Slicing \\ + paths [s]} & \makecell{Optim. [s]} & Total [s] & \makecell{Curved slicing \\ preparation [s]} & \makecell{Slicing \\ + paths [s]} & Total [s] & \makecell{Curved slicing \\ preparation [s]} & \makecell{Slicing \\ + paths [s]} & Total [s] \\
\hline
Car model & 331.2 & 0.6 & 493.7 & \textbf{\textcolor{red}{871.7}} & 6.7 & 0.5 & \textbf{\textcolor{green}{19.2}} & 7.1 & 0.5 & 38.8 \\
Dome & 66.8 & 0.5 & 363.3 & \textbf{\textcolor{red}{436.9}} & 3.1 & 0.5 & \textbf{\textcolor{green}{25.6}} & 3.0 & 0.4 & 38.2 \\
Robotic ankle & 62.5 & 0.6 & 46.3 & 118.9 & 27.3 & 0.5 & \textbf{\textcolor{green}{69.6}} & 31.0 & 0.4 & \textbf{\textcolor{red}{423.8}} \\
Jet & 186.8 & N/A & N/A & N/A & 57.8 & 0.5 & \textbf{\textcolor{green}{84.8}} & 78.6 & 0.4 & 269.8 \\
Insole & 74.8 & N/A & N/A & N/A & 7.4 & 0.4 & \textbf{\textcolor{green}{28.1}} & 7.4 & 0.5 & 100.1 \\
Cheese stopper & 295.7 & 0.6 & 1612.9 & \textbf{\textcolor{red}{1942.8}} & 7.4 & 0.4 & \textbf{\textcolor{green}{51.8}} & 6.5 & 0.4 & 153.3 \\
Wing & 19.8 & 0.6 & 4.9 & 31.0 & 2.2 & 0.4 & \textbf{\textcolor{green}{23.1}} & 2.1 & 0.4 & \textbf{\textcolor{red}{274.3}}\\
Eggs box & 595.0 & 0.6 & 698.2 & 1347.3 & 11.5 & 0.4 & \textbf{\textcolor{green}{86.5}} & 12.1 & 0.5 & \textbf{\textcolor{red}{1307.4}}\\
Foil cutter & 40.9 & 0.5 & 400.9 & \textbf{\textcolor{red}{448.5}} & 5.1 & 0.4 & \textbf{\textcolor{green}{26.3}} & 5.0 & 0.5 & 55.3 \\
Billinge & 355.6 & N/A & N/A & N/A & 2.2 & 0.5 & \textbf{\textcolor{green}{34.1}} & 3.2 & 0.5 & 60.0 \\
Link's shield & N/A & N/A & N/A & N/A & 45.0 & 0.5 & \textbf{\textcolor{green}{92.1}} & 49.3 & 0.5 & 224.8 \\
\hline
\end{tabu}
\caption{This table shows a time comparison between Curvislicer~\cite{etienne_19_curvislicer} and our method. For our method we measure two versions: standard toolpath generation and our orientation optimized paths (\textit{optizor}). We breakout the different steps of the algorithms, and highlight in green the fastest times and red the slowest.}
\label{tab:times}
\end{table*}
%

Ahlers et al.~\cite{ahlers_19_3d} algorithm is extremely fast as it performs no optimization. However, it is less general as shown in Figure~\ref{fig:two_towers}. In addition, on several of our tests the implementation was not able to produce any curved layers, in particular the insole and jet models. While this may be due to an implementation issue, it is also the case that any collision will discard an entire surface from being curved. This quickly becomes limiting on complex surfaces.


\section{Conclusions}
\label{sec:limitsandfuture}

We propose a new approach which explores a different tradeoff between generality and computational efficiency. It is faster and simpler to implement than optimizing for a volume deformation~\cite{etienne_19_curvislicer} while being more general than more direct approaches curving fully accessible top surfaces~\cite{ahlers_19_3d}.

In addition we introduce two novel ingredients: 1) A filter to remove spurious tiny features within otherwise smoothly curved regions. This is especially useful on scanned or natural surfaces. 2) A path orientation strategy that allows to choose either to align along the maximum principal curvature, or the minimal one (or any angle in between, in fact). Aligning with the maximum allows to avoid a staircase defect due to the paths being across the curvature. Aligning with the minimal allows a more natural transition between curved parts and planar parts of a print.

\paragraph*{Limitations and future work.} Our technique has a number of drawbacks, that are also venues of future work.

In terms of generality, we can only consider the top surface, while the volume approach in~\cite{etienne_19_curvislicer} can curve surfaces everywhere in the part.

In terms of optimization, the path orientation optimizer is currently called at every slice to obtain solutions having different phases (and hence alignments). This incurs a cost that is probably only justified on the visible layers. Therefore, we plan to modify our slicer to use standard zigzag infills inside and optimize the orientations only on top covers.

We believe our technique could be recast in the context of multi-axis robotic platforms, where the nozzle can incline. The gouging issue is less problematic in this context and the orientation of paths within layers could be further optimized for structural requirements~\cite{fang_20_reinforced,zhang_22_s3}.


\bibliographystyle{plain}
\bibliography{biblio}

\end{document}